\documentclass{aa}
\usepackage[english]{babel}
\usepackage[utf8x]{inputenc}
\usepackage{graphicx}
\usepackage{txfonts}
\usepackage{natbib}
\usepackage{subfigure}
\usepackage{arydshln}
\usepackage{changepage}
\usepackage{afterpage}
\usepackage{float}
\usepackage{ulem}
\usepackage[dvipsnames]{xcolor}
\usepackage[colorinlistoftodos]{todonotes}
\usepackage[colorlinks=true, urlcolor=black, citecolor=blue, linkcolor=red]{hyperref}

\begin{document}

\newcommand{\unit}[1]{\; \ensuremath{\mathrm{#1}\xspace}} 
\newcommand{\GeV}{\unit{GeV}}
\newcommand{\MeV}{\unit{MeV}}
\newcommand\shorttitle{}

\title{Disentangling hadronic from leptonic emission in the composite SNR G326.3$-$1.8}
\author{
J.~Devin$^{(1)}$ \and 
F.~Acero$^{(2)}$ \and 
J.~Ballet$^{(2)}$ \and 
J.~Schmid$^{(2)}$
}

\titlerunning{Disentangling hadronic from leptonic emission in G326.3$-$1.8}
\authorrunning{J. Devin et al.}

\institute{
\inst{1}~Laboratoire Univers et Particules de Montpellier, Universit\'e Montpellier, CNRS/IN2P3, F-34095 Montpellier, France\\
\email{jdevin.phys@gmail.com} \\
\inst{2}~Laboratoire AIM, CEA-IRFU/CNRS/Universit\'e Paris Diderot, Service d'Astrophysique, CEA Saclay, F-91191 Gif sur Yvette, France\\
\email{fabio.acero@cea.fr} \\
}
	\date{Received: 13 March 2018 / Accepted: 25 May 2018}

\abstract
{G326.3$-$1.8 (also known as MSH 15$-$5\textit{6}) has been detected in radio as a middle-aged composite supernova remnant (SNR) consisting of an SNR shell and a pulsar wind nebula (PWN), which has been crushed by the SNR's reverse shock. Previous $\gamma$-ray studies of SNR G326.3$-$1.8 revealed bright and extended emission with uncertain origin. Understanding the nature of the $\gamma$-ray emission allows probing the population of high-energy particles (leptons or hadrons) but can be challenging for sources of small angular extent.}
{With the recent $\textit{Fermi}$ Large Area Telescope data release Pass 8 providing increased acceptance and angular resolution, we investigate the morphology of this SNR to disentangle the PWN from the SNR contribution. In particular, we take advantage of the new possibility to filter events based on their angular reconstruction quality.  }
{We perform a morphological and spectral analysis from 300~MeV to  300 GeV. We use the reconstructed events with the best angular resolution (PSF3 event type) to separately investigate the PWN and the SNR emissions, which is crucial to accurately determine the spectral properties of G326.3$-$1.8 and understand its nature.}
{The centroid of the $\gamma$-ray emission evolves with energy and is spatially coincident with the radio PWN at high energies (E $>$ 3 GeV). The morphological analysis reveals that a model considering two contributions from the SNR and the PWN reproduces the $\gamma$-ray data better than a single-component model. The associated spectral analysis using power laws shows two distinct spectral features, a softer spectrum for the remnant ($\Gamma$ = 2.17 $\pm$ 0.06) and a harder spectrum for the PWN ($\Gamma$ = 1.79 $\pm$ 0.12), consistent with hadronic and leptonic origin for the SNR and the PWN respectively. Focusing on the SNR spectrum, we use one-zone models to derive some physical properties and, in particular, we find that the emission is best explained with a hadronic scenario in which the large target density is provided by radiative shocks in H\,{\sc i} clouds struck by the SNR.}
{}


\maketitle


\section{Introduction}
Supernova remnants (SNRs) and pulsar wind nebulae (PWNe) have long been considered potential sources of Galactic cosmic rays and have therefore been investigated over a wide range of energies. In SNRs, the fast shock wave propagating into the interstellar medium (ISM) or the circumstellar medium is thought to accelerate particles (electrons and protons), which gain energy through first order Fermi acceleration \citep{Bell:1978} also known as the Diffusive Shock Acceleration mechanism (DSA). In core-collapse SNRs, a very fast rotating and highly magnetized pulsar can give rise to a PWN, in which electrons and positrons from the pulsar wind are re-accelerated to relativistic energies at a termination shock. These Galactic accelerators have mostly been studied independently while in case of core-collapse SNRs, the SNR, the PWN and the pulsar are part of the same object. However, for systems with angular sizes smaller or comparable to the instrument point-spread function (PSF), it can be difficult to assess the origin of the emission, in particular at $\gamma$-ray energies where the angular resolution is comparatively much larger than in the radio and X-ray ranges. Although it can be challenging to understand their origin, these $\gamma$ rays allow probing the population of high-energy particles, such as accelerated electrons interacting with the Cosmic Microwave Background (CMB) or other target photons by Inverse Compton (IC) scattering, and also accelerated protons interacting with gas that produce neutral pions which decay into $\gamma$ rays.
Morphological studies complementing purely spectral analyses have the potential to help identify multiple particle acceleration regions in one object such as interaction regions with surrounding clouds or the different emissions coming from the SNR and/or the PWN in composite objects. 

With an SNR shell and a PWN seen at radio wavelengths, the Galactic SNR G326.3$-$1.8  is a prototype of the so-called composite SNRs \citep{Mills, Milne}. Its distance is estimated between 3.1 kpc \citep{Goss:1972} and 4.1 kpc \citep{Rosado:1996} as established by the H\,{\sc i}  absorption profile and H$\alpha$ velocity measurements respectively. \citet{Temim:2013} estimated this SNR to be 16,500 years old with a shock velocity of 500 km s$^{-1}$, expanding in an ISM density of $n_{0}$ = 0.1 cm$^{-3}$. 
\begin{figure*}[htp!]
\begin{minipage}[b]{.48\linewidth}
  \centering
  \includegraphics[scale=0.43]{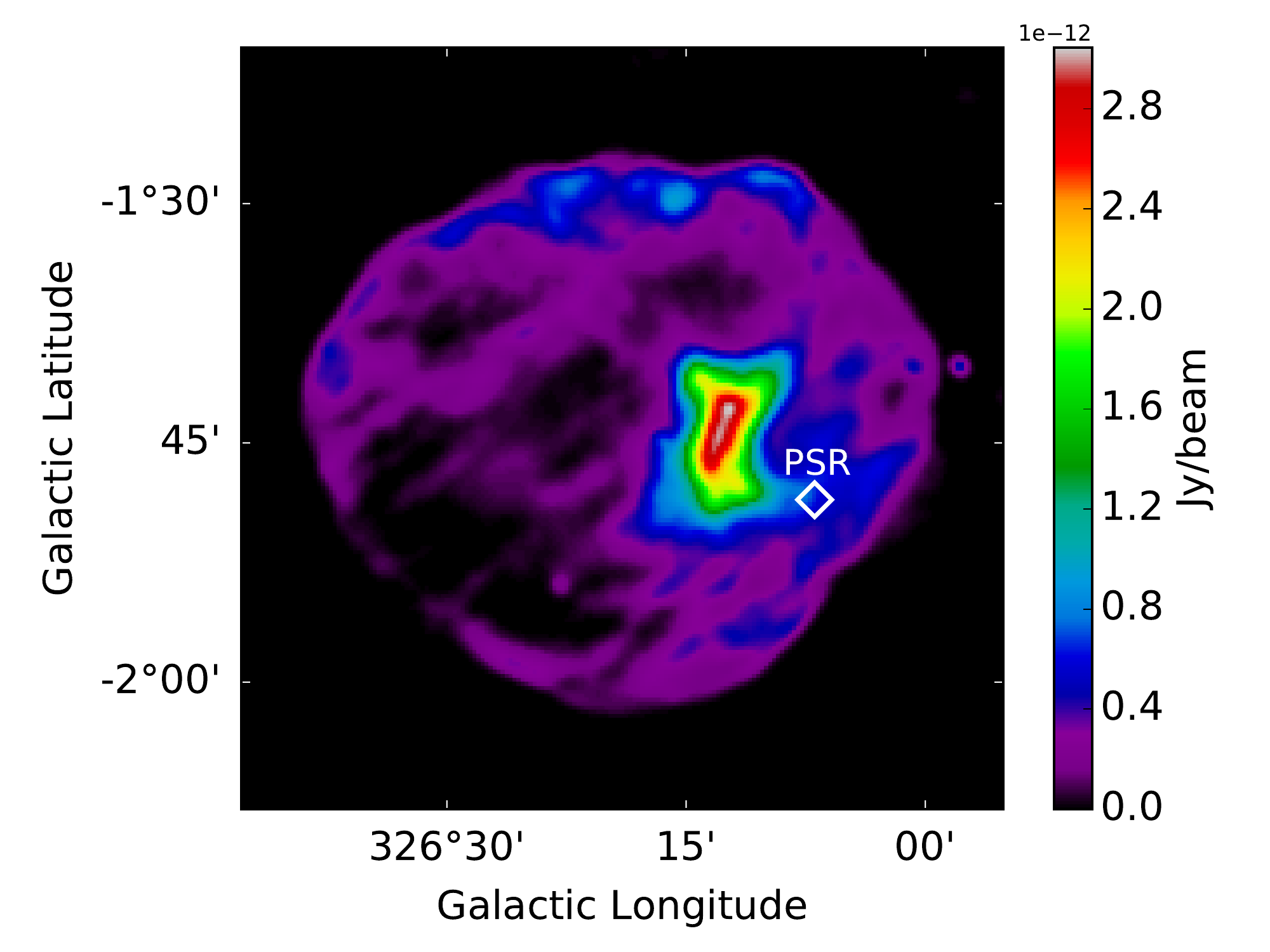}
  \caption{\small 843 MHz MOST radio image of SNR G326.3$-$1.8 \citep{MOSTcat:1996}.  The position of the pulsar candidate is represented by a white diamond. The positional uncertainty is much smaller than the marker size. \label{fig:MOST}}
 \end{minipage} \hfill
 \begin{minipage}[b]{.48\linewidth}
  \centering
  \includegraphics[scale=0.43]{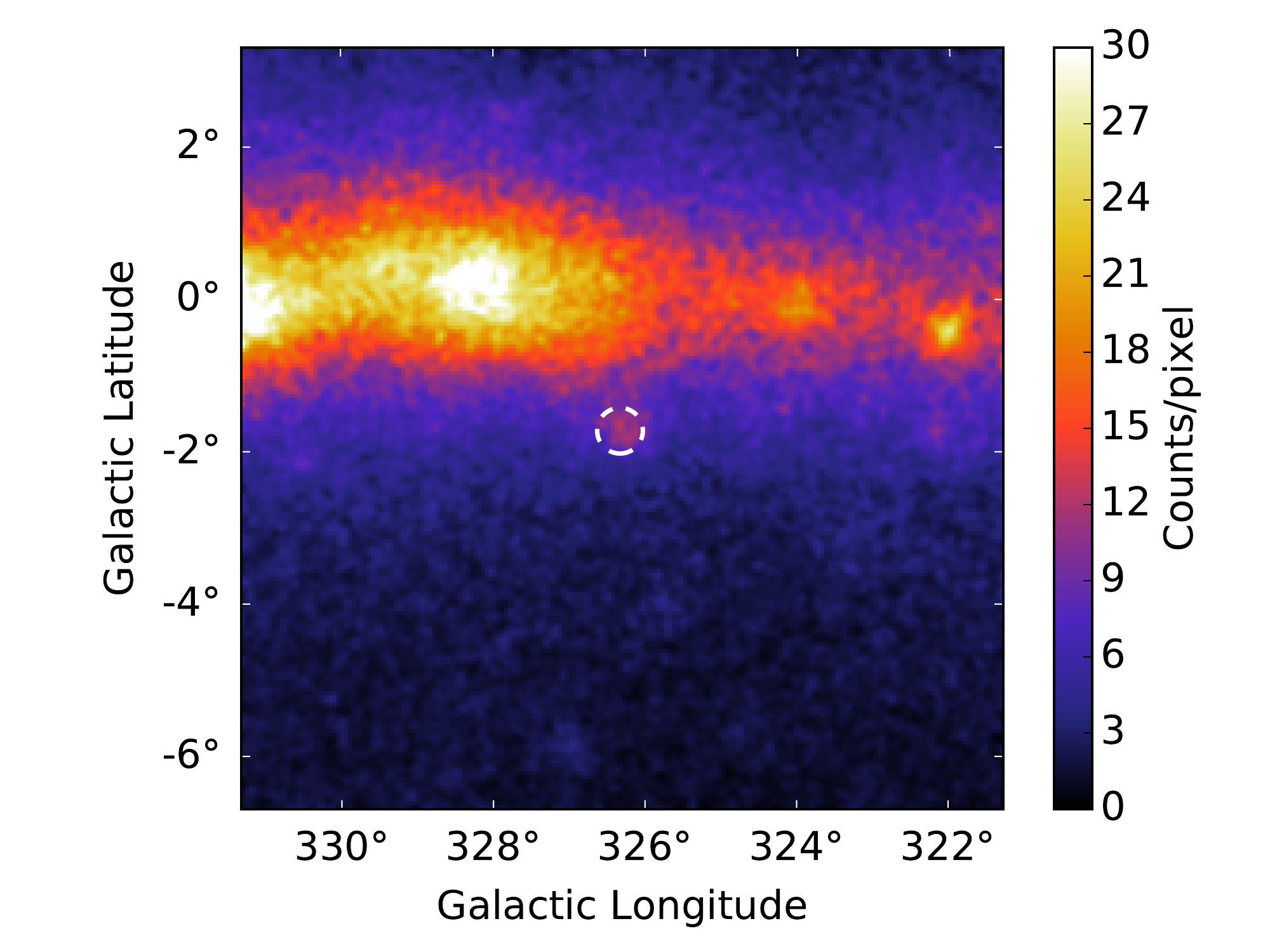}
  \caption{\small Counts map from 300 MeV to 300 GeV of a 10° $\times$ 10° region centered on the position of the SNR (dashed circle) and using the PSF3 events. The pixel size is 0.05°.\label{fig:Cmap}}
  \vspace{0.3cm}
 \end{minipage}
\end{figure*}
Figure~\ref{fig:MOST}, obtained from radio observations \citep{MOSTcat:1996}, shows a symmetric SNR shell with 0.3° radius and a PWN trailing the putative pulsar and likely crushed by the remnant's reverse shock. Non-thermal radio emission has been reported with a spectral index of $\alpha$ = 0.34 for the shell and $\alpha$ = 0.18 for the nebula where $S_{\nu}$ $\propto$ $\nu^{-\alpha}$ \citep{Dickel:2000}. Optical H$\alpha$ filaments were observed in the southwest and northeast parts of the remnant, and appear to spatially correlate with the shell \citep{Vandenbergh:1979,Dennefeld:1980} indicating the presence of neutral material at the shock front. The PWN component is highly polarized with a luminosity of L($10^{7} − 10^{11}$ Hz) ∼ $5 × 10^{34}$ erg s$^{−1}$ \citep{Dickel:2000}. The associated pulsar has not been detected but \textit{Chandra} maps have revealed a point source embedded in the X-ray PWN located in the southwest of the radio nebula \citep{Temim:2013}. SNR G326.3$-$1.8 was also detected in X-rays by \textit{ROSAT} \citep{Kassim:1993} and \textit{ASCA} \citep{Plucinsky:1998} showing a complete shell that spatially correlates with the radio SNR while the width of the PWN in X-rays shrinks near the compact object.
At higher energies, previous $\gamma$-ray studies have revealed emission with uncertain origin \citep{Temim:2013} and SNR G326.3$-$1.8 has only recently been found to be extended with the $\textit{Fermi}$-LAT data \citep{First_SNRcat:2016}. 

The latest Large Area Telescope data release Pass 8 \citep{Pass8:Atwood:2013} allows not only a claim of significant extension of the $\gamma$-ray emission, but also a study of the PWN and SNR contributions separately.
This distinction might be crucial for understanding the underlying emission mechanisms and potentially distinguishing between hadronic and leptonic nature of the constituents. 
In this paper, we briefly describe the latest data release Pass 8 before presenting a morphological study of SNR G326.3$-$1.8. In particular, we investigate its energy-dependent morphology and model the emission with different templates. We also report a spectral analysis of our best models using two spatial components for the $\gamma$-ray emission and derive physical properties using one-zone models for the SNR spectrum. 

\section{\textit{Fermi}-LAT and Pass 8 description \label{sec:Fermi}}

The Large Area Telescope (LAT) on board the $\textit{Fermi}$ satellite is a pair-conversion instrument sensitive to $\gamma$ rays in the energy range from 30 MeV to more than 300 GeV. \\
Since the launch in August 2008, the $\textit{Fermi}$-LAT event reconstruction algorithm has been progressively upgraded to make use of the increasing understanding of the instrument performance as well as the environment in which it operates.
Following Pass 7, released in August 2011, Pass 8 is the latest version of the $\textit{Fermi}$-LAT data release \citep{Pass8:Atwood:2013}. The enhanced reconstruction and classification algorithms result in improvements of the effective area, the PSF and the energy resolution. 
One major advance with respect to previous releases is the classification of detected photon events according to their reconstruction quality. The data set is hence divided into types of events with different energy or angular reconstruction qualities.
The PSF selection divides the data into four parts: from PSF0 to PSF3, the latter being the quartile with the best angular resolution (68\% containment radius of 0.4° at 1 GeV compared to 0.8° without selection). 
This type of event selection, combined with the large amount of data collected by the LAT since its launch, makes Pass 8 $\gamma$ rays a powerful tool to identify and study extended $\gamma$-ray sources.

\section{Data analysis\label{sec:ata}}

We perform a binned analysis using 6.5 years of data collected from August 4, 2008 to January 31, 2015, within a 10° $\times$ 10° region around the position of SNR G326.3$-$1.8. Since the object remains significant with $25\%$ of the data (more than 24$\sigma$ between 300 MeV and 300 GeV), we take advantage of the new PSF3 selection to limit contamination between the PWN and the SNR components as well as that from the Galactic plane.
We select events between 300 MeV and 300 GeV, with a maximum zenith angle of 100° to reduce the contamination of the bright Earth limb. 
Time intervals during which the rocking angle  of the satellite was more than 52° are excluded as well as those during which it passed through the South Atlantic Anomaly. We set the pixel size to 0.05° and divide the whole energy range (300 MeV -- 300 GeV) into 30 bins. 
We use version 10 of the \verb|Science Tools| (v10r0p5) and the \verb|P8R2_V6| Instrument Response Functions (IRFs) with the \verb|SOURCE| event class for the following analysis\footnote{The Science Tools package and related documentation are distributed by the \textit{Fermi} Science Support Center at https://fermi.gsfc.nasa.gov/ssc}. 
The resulting count map of the 10° $\times$ 10° region centered on the position of the SNR is shown in Figure~\ref{fig:Cmap}.

The $\gamma$-ray data around the source are modeled starting with the $\textit{Fermi}$-LAT 3FGL source catalog \citep{3FGL}, complemented by the extended source FGES J1553.8$-$5325\footnote{From the \textit{Fermi} Galactic Extended Source catalog} \citep{FGES:Pass8}.
We first fit the point sources and extended sources within a 10° radius (additionally accounting for the most significant sources between 10° and 15°) simultaneously with the Galactic and isotropic diffuse emissions described by the files \verb|gll_iem_v06.fits| \citep{GIEM_Fermi:2016} and \verb|iso_P8R2_SOURCE_V6_v06_PSF3.txt| respectively \footnote{Available at \url{https://fermi.gsfc.nasa.gov/ssc/data/access/lat/BackgroundModels.html}}. We then compute a residual test statistic (TS) map to search for additional sources. 

\begin{figure}
\centering
  \includegraphics[scale=0.45]{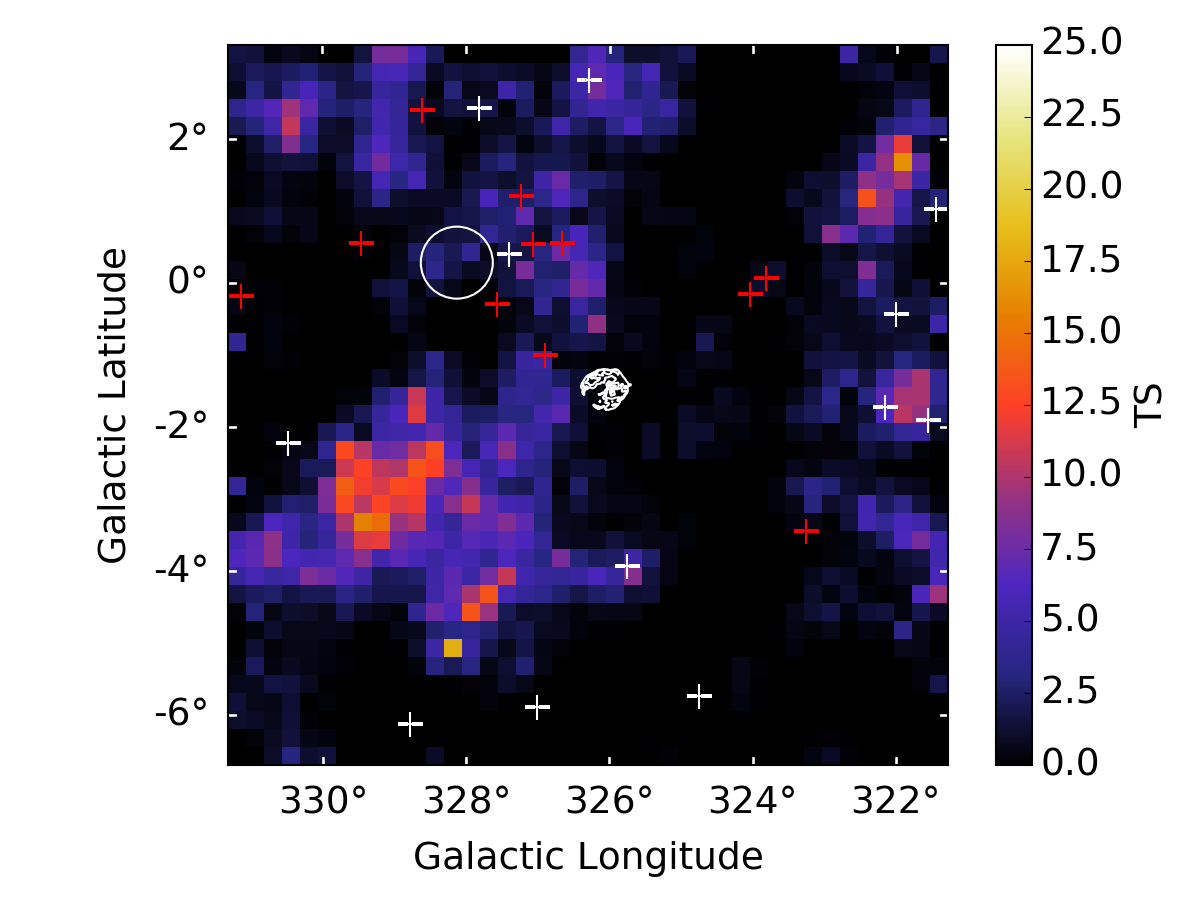}
  \caption{\small Residual TS map from 300 MeV to 300 GeV of a 10° $\times$ 10° region centered on the SNR and using the PSF3 events. The pixel size is 0.25° and the radio contours of the SNR are overlaid in white. The white circle is an FGES extended source. The white crosses are the 3FGL point sources and the red crosses are the sources added to the model. \label{fig:TSmap_all_sky}}
\end{figure}

The TS is defined to test the likelihood of one hypothesis $\mathcal{L}_1$ (including a source) against the null hypothesis $\mathcal{L}_0$ (absence of source),  such that:
\begin{equation} 
\hspace{2.5cm} \textrm{TS} = 2 \times (\log{\mathcal{L}_1 } - \log{\mathcal{L}_0})  \label{eq:TS}  
\end{equation}
This can be directly interpreted in terms of significance of hypothesis $1$ with respect to the null hypothesis $0$ in which the TS follows a $\chi^2$-law with $n$ degrees of freedom for $n$ additional parameters.
To evaluate the significance of putative new sources, we compute a two-dimensional residual TS map that tests the hypothesis of a point source with a generic $E^{-2}$ spectrum against the null hypothesis at each point in the sky. 
The positions where the TS values exceed 25 (corresponding to a significance of more than $4\sigma$) are used as seeds to identify  $\gamma$-ray sources in addition to the 3FGL. In that way, we iteratively add eleven sources in the 10° $\times$ 10° region and we fix their spectral parameters to their best-fit values found with \verb|gtlike|.  Figure~\ref{fig:TSmap_all_sky} shows the final residual TS map including all the sources. Note that the apparent diffuse residual emission (for which TS$_{\rm{max}}$ $\approx$ 17) disappears above 500 MeV.

\subsection{Morphological analysis\label{sec:Morphological_analysis}}

\subsubsection{Extension\label{sec:Extension}}

The 3FGL catalog compiled by the $\textit{Fermi}$-LAT collaboration \citep{3FGL} has two point-like sources tentatively associated with the SNR, which we remove for our analysis. Since the creation of the first SNR catalog \citep{First_SNRcat:2016}, G326.3$-$1.8 has been known to show extended $\gamma$-ray emission, and its radius has been determined to be $0.21^\circ$ using an extended uniform disk model, somewhat smaller than the $0.31^\circ$ radius of the radio shell but larger than the radio PWN.
However, that analysis was based on only three years of data and made use of the former Pass 7 data release. 

\begin{figure*}[th!]
\centering
\includegraphics[width=0.4\textwidth]
{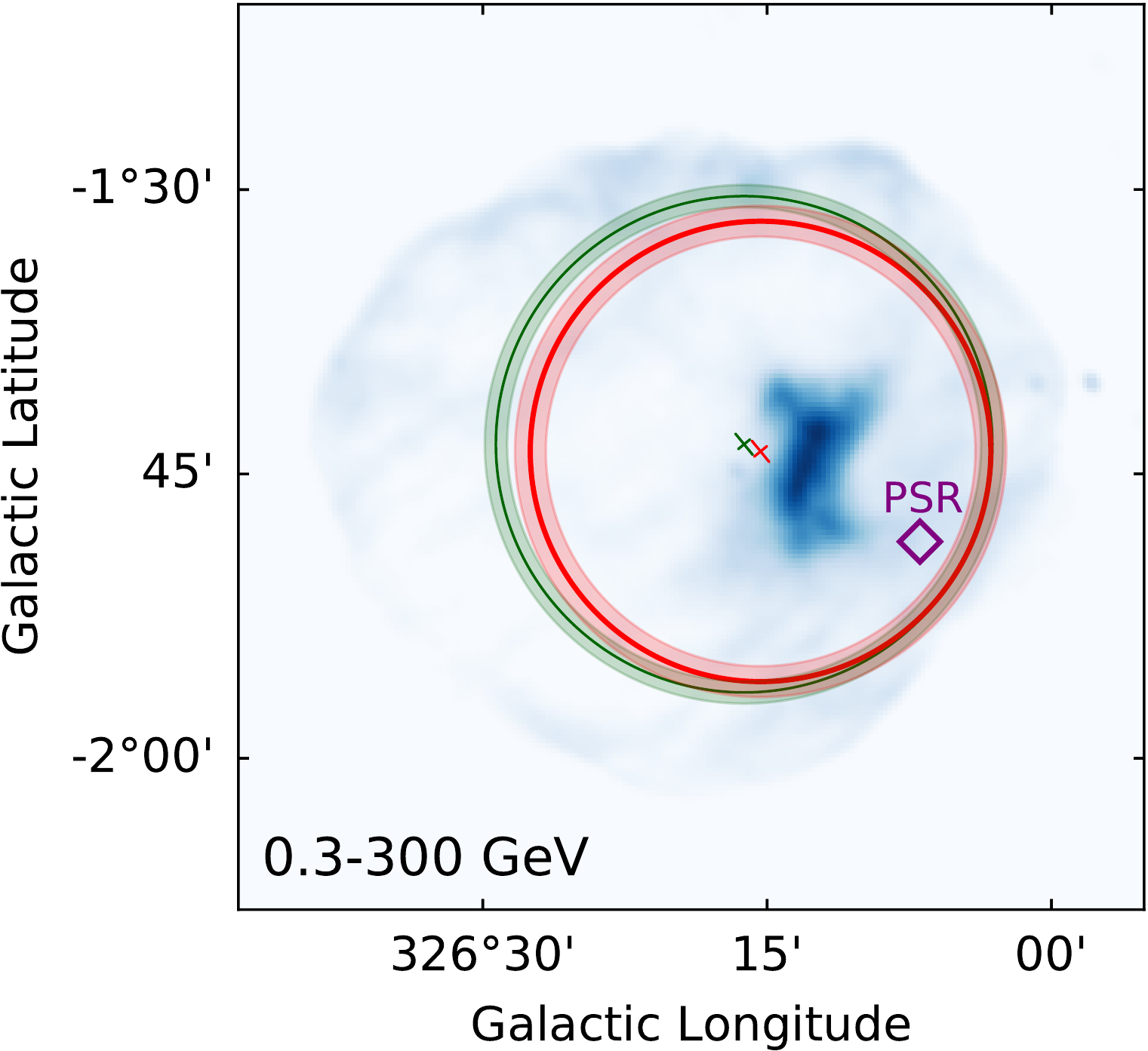}
\hspace{1.5cm}
\includegraphics[width=0.4\textwidth]{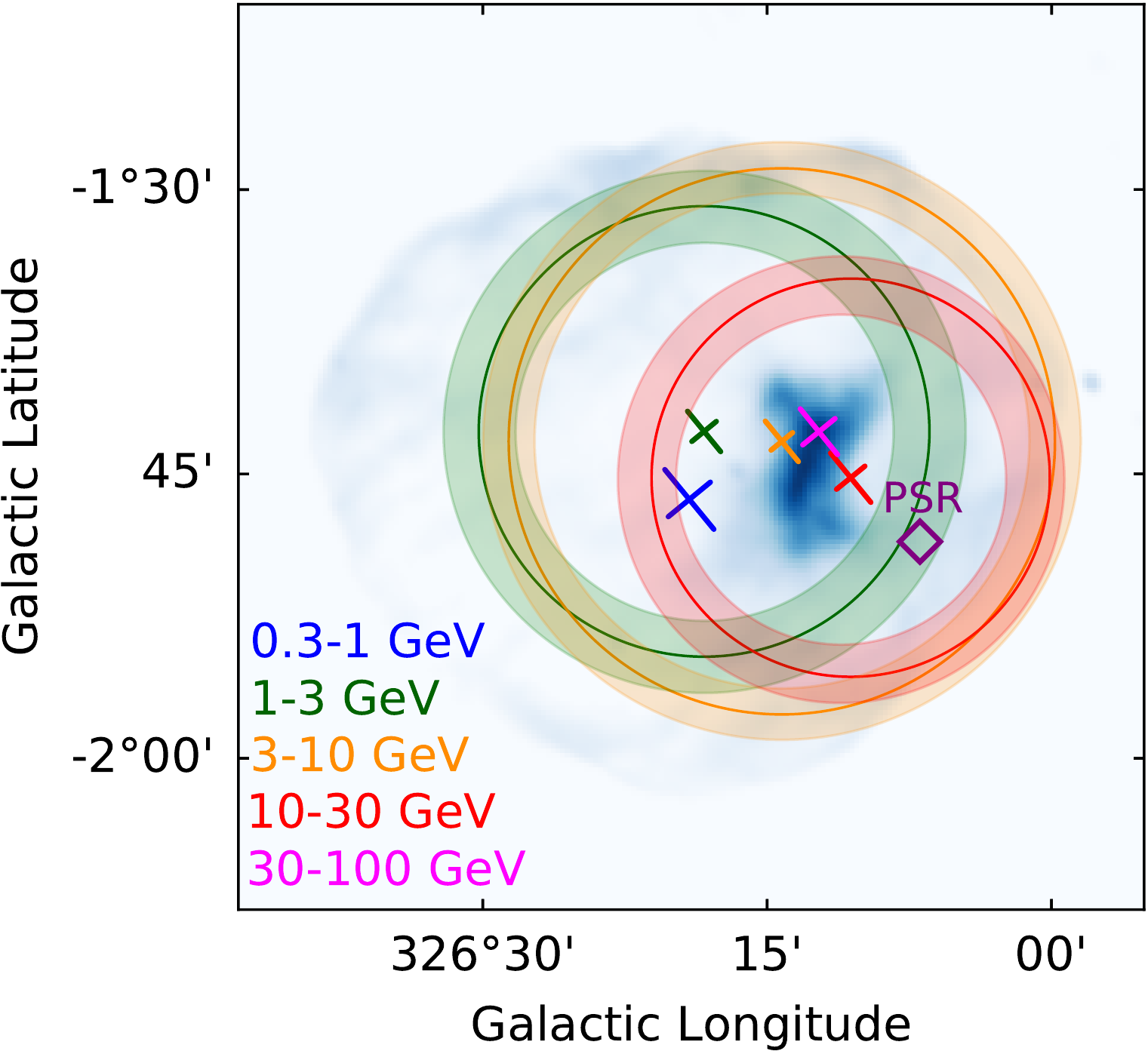}
\caption{\label{fig:Best_Disk_Gaussian} \textit{(Left)} \small Best-fit extended source from 300 MeV to 300 GeV for a uniform disk hypothesis (green) and a 2D-symmetric Gaussian (red), plotted on the MOST radio image. The crosses are the centroid uncertainties (1$\sigma$), the solid circles correspond to the $r_{68}$ of the uniform disk and the Gaussian template and the shaded areas are the 1-$\sigma$ error on size. The putative position of the pulsar is represented by a diamond. \textit{(Right)} Best-fit Gaussian or point source in five energy bands. The crosses are the centroid and the position uncertainties (1$\sigma$). If the source is significantly extended (see Figure~\ref{fig:TS_and_TSext}), the $r_{68}$ of the Gaussian is depicted by a solid circle with the associated errors (shaded areas), otherwise the best-fit point source is represented.}
\end{figure*}

With the latest Pass 8 data and using the PSF3 event type, we revisit the morphology of this SNR to understand the nature of the $\gamma$-ray emission. We start by finding the best position of a point source, modeling its emission as a power law and using the \verb|pointlike| framework \citep{Kerr:2010} from 300 MeV to 300 GeV. Then, we investigate the extension of the $\gamma$-ray emission using a 2D-symmetric Gaussian and a disk with a uniform brightness. Table~\ref{tab:Fit_Pointlike} shows the respective best-fit position and extension -- if extended -- for the different spatial models.
The significance of a source extension is expressed in terms of the test statistic  TS$_{\rm ext}$, where the hypothesis of the best-fit extended spatial model is tested against the null hypothesis of the best-fit point-like source. Given that in both hypotheses the localization of the source is optimized, the extended source model adds one degree of freedom -- the source size -- with respect to the point-source model. 
Thus, the significance can be directly interpreted as the square root  $\sqrt{\rm{TS_{ext}}}$.  As reported in Table~\ref{tab:Fit_Pointlike}, the $\gamma$-ray emission is extended with more than 13$\sigma$ confidence level and the uniform disk radius is found to be $r$ = $0.266^\circ$ $\pm$ 0.012$^\circ$. Figure~\ref{fig:Best_Disk_Gaussian} (left) shows the best-fit position and extension (68\% containment radius $r_{68}$) for the disk and the Gaussian, plotted on the radio image, with the associated uncertainties. The centroid of each extended model is slightly shifted toward the radio PWN but is not coincident with its position.  No significant residual emission appears in the residual TS map (not shown here) including either the disk or the Gaussian in the model.

\begin{table*}[t!]
	\centering
	\begin{tabular}{l|ccccccc}
		\hline
        \hline
        \rule{0pt}{2.ex} 
        \hspace{0cm}
        Spatial model & RA$_{\rm{J2000}}$ ($^\circ$)& Dec$_{\rm{J2000}}$ ($^\circ$) &  $r$ or $\sigma$ ($^\circ$) & $r_{68}$ ($^\circ$) & TS & TS$_{\rm ext}$\\ \hline
        \rule{0pt}{2.ex} 
        \hspace{0cm}
        Point source & 238.167 $\pm$ 0.009 & $-$56.181 $\pm$ 0.008 & --- & --- & 689.5 & ---\\
        \rule{0pt}{2.ex} 
        \hspace{0cm}
        Disk & 238.170 $\pm$ 0.012 & $-$56.152 $\pm$ 0.012 & 0.266 $\pm$ 0.012 & 0.218 $\pm$ 0.010 & 866.5 & 177.0 \\
        \rule{0pt}{2.ex} 
        \hspace{0cm}
        Gaussian & 238.157 $\pm$ 0.013 & $-$56.166 $\pm$ 0.012 & 0.134 $\pm$ 0.009 & 0.202 $\pm$ 0.014 & 863.4 & 173.9 \\ \hline
	\end{tabular}
   	\caption{\label{tab:Fit_Pointlike} \small Best-fit positions and sizes (radius or sigma) with the associated statistical errors using different spatial models. $r_{68}$ corresponds to the 68\% containment radius of each extended model. The TS and TS$_{\rm ext}$ values are also given.} 
\end{table*}

\subsubsection{Energy-dependent morphology\label{sec:E_dpd_morpho}}

\begin{figure}[th!]
\includegraphics[scale=0.29]{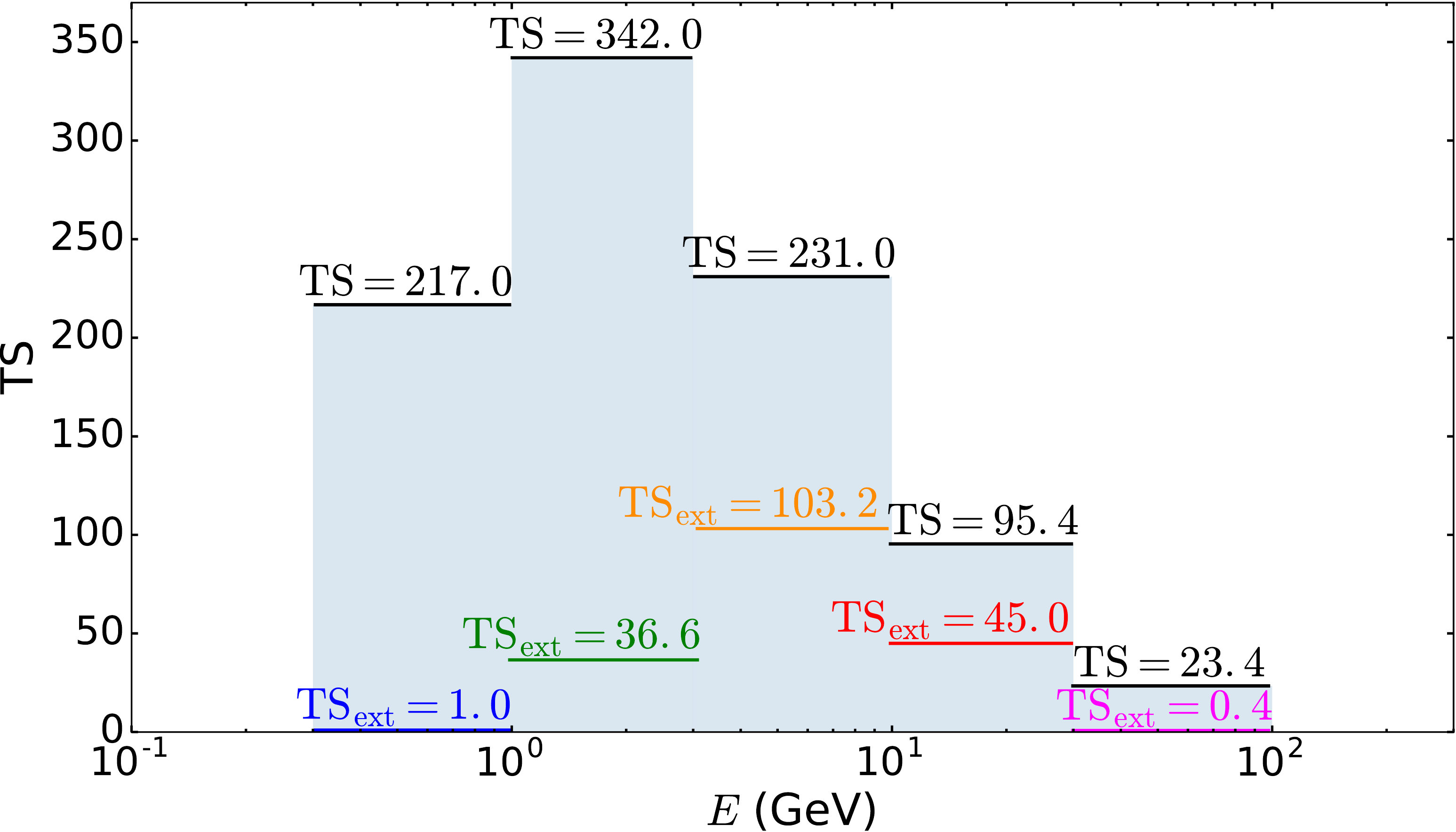}
\caption{\label{fig:TS_and_TSext}\small Test statistic of the source (black bars) and of extension (colored bars) for the best-fit Gaussian in individual energy bands.}
\end{figure}

Although the $\gamma$-ray emission can be adequately described with a one-component model, either a disk or a 2D-symmetric Gaussian, this stands in slight tension to the discovery of a hard point-like $\gamma$-ray source above 50 GeV \citep{2FHL} at the location of the PWN, clearly displaced from the center of the SNR shell. To investigate the morphology in more detail, we divide the data into five logarithmically spaced energy bins from 300 MeV to 300 GeV that we subsequently fit individually with \verb|pointlike| using a 2D-symmetric Gaussian. Because the PSF width depends strongly on energy up to $\sim$ 10~GeV, and our energy bins are quite broad (half a decade), we need to adopt a specific spectral model for the source. We describe it with a power law with free spectral index. The normalizations of the source, the Galactic and isotropic diffuse emissions are let free while the spectral parameters of the other sources are fixed to their best-fit values.
Figure~\ref{fig:Best_Disk_Gaussian} (right) depicts  the results of the fitting procedure in the individual energy bands. At low energies (300 MeV -- 1 GeV), the PSF ($r_{68}$ $\sim$ 0.4$^{\circ}$ at 1 GeV) is larger than the SNR radius. An extended source of the SNR size is compatible with the data but is not significantly better than a point source. The associated best-fit position lies outside the radio PWN. Between $1$ and $3$ GeV, the significance of the extension is more than 5$\sigma$ (the values are reported in Figure~\ref{fig:TS_and_TSext}) and the position of the Gaussian appears to be fairly consistent with the center of the radio SNR. At higher energies (from 3 to 30 GeV), the $\gamma$-ray morphology is still significantly extended (more than 5$\sigma$) and the centroid of the best-fit Gaussian gets closer to the radio PWN. Above 30 GeV, the $\gamma$-ray emission is not significantly extended and the best-fit position lies inside the radio PWN.

\subsubsection{Building a more detailed model}

This energy-dependent source morphology clearly requires a more detailed investigation beyond a one-component modeling. Since the PSF below 1 GeV is not small enough to resolve the SNR, the following morphological analysis uses data between 1 and 300 GeV.

\begin{figure*}[!th]
\centering
\includegraphics[scale=0.70]{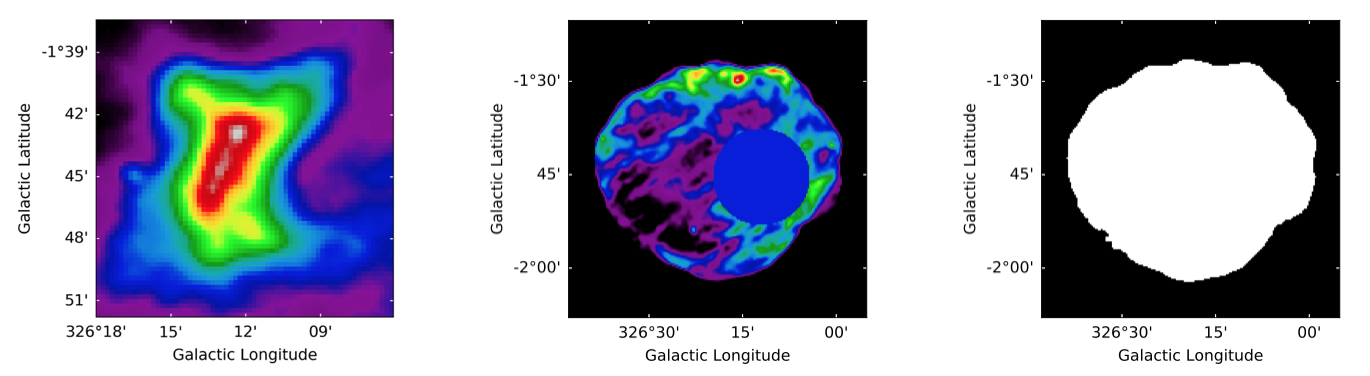}
\caption{\small Templates used in this analysis: \textit{(left)} the radio PWN, \textit{(center)} the radio SNR with the PWN contribution removed and filled with the average value around it (blue disk), \textit{(right)} the SNR mask derived from the SNR radio template and filled homogeneously. The radio PWN \textit{(left)} does not have the same scale as the two other templates.\label{fig:Templates}}
\end{figure*}

Electrons and positrons, accelerated in the PWN, that radiate by synchrotron emission are expected to also radiate in the GeV band by IC scattering on photon fields. Since this SNR is relatively young, particles are well-confined inside the PWN. \cite{Temim:2013} estimate the magnetic field to be $B_{\rm{PWN}}$ $\approx$ 34 $\mu$G when the SNR has already begun significantly compressing the PWN at an age of 19,000 years. The emission seen in the radio band should track the older accelerated electrons and we expect that the extension of the $\gamma$-ray emission should not be larger than the radio emission. We thus model the $\gamma$-ray emission from the PWN using its radio template (see Figure~\ref{fig:Templates}, left panel), knowing that the magnetic field spatial distribution inside the PWN should only moderately impact our model given the small size of the PWN compared to the $\textit{Fermi}$-LAT PSF.

\begin{figure*}[th!]
  \includegraphics[width=0.5\textwidth, scale=0.4]{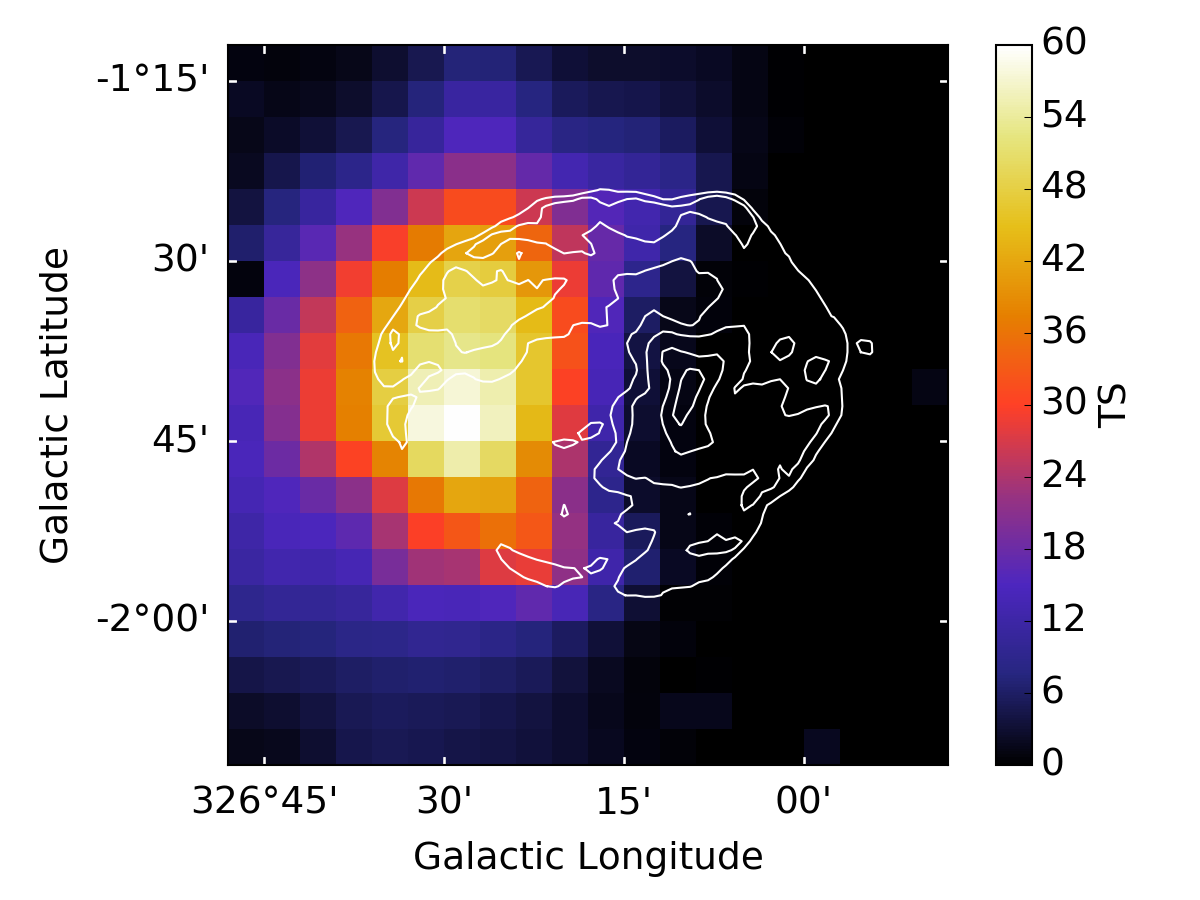}
  \includegraphics[width=0.5\textwidth, scale=0.4]{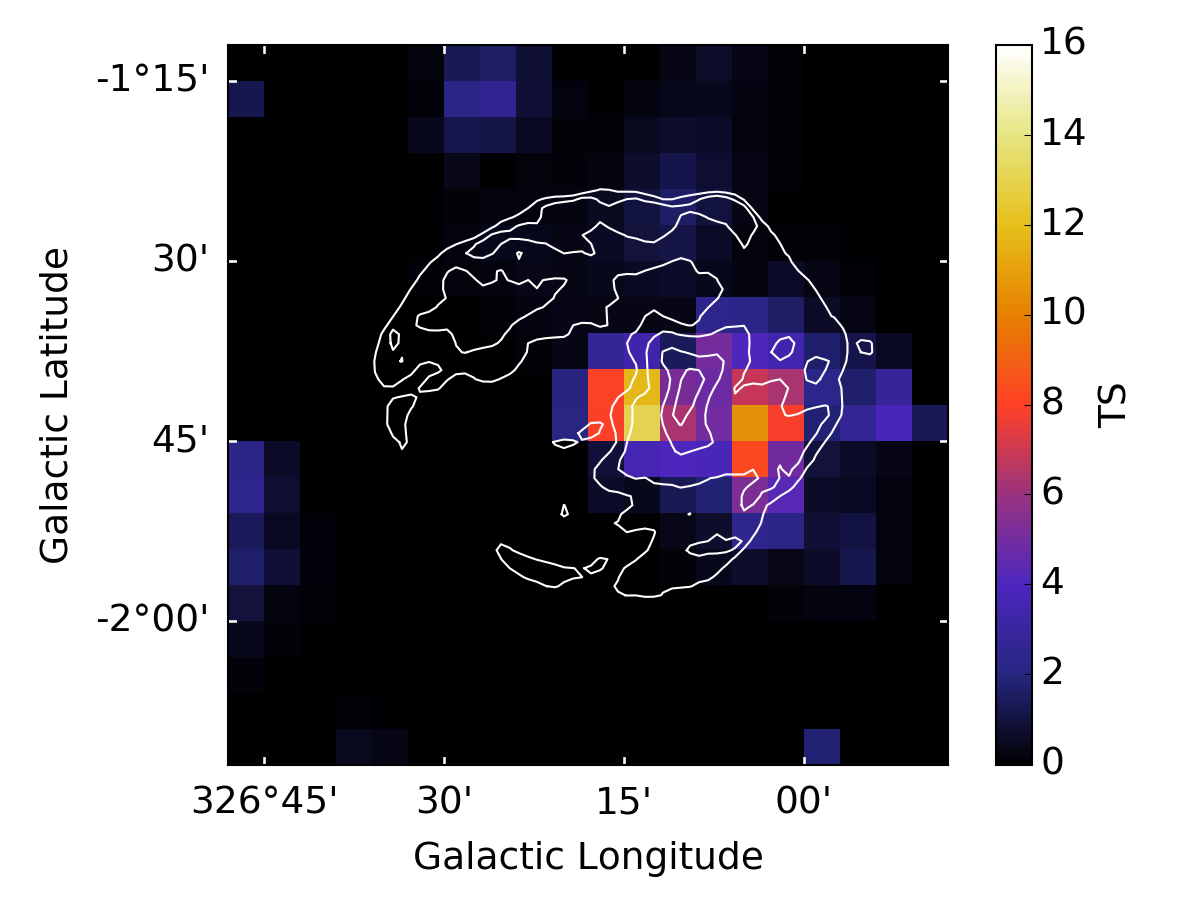}
  \caption{\small Residual 1° $\times$ 1° TS maps from 1 GeV to 300 GeV when we use only the radio PWN (\textit{left}) \label{fig:TSmap_PWN} or the SNR mask (\textit{right}) to describe the $\gamma$-ray emission (note that the TS bars do not have the same scale). The pixel size is 0.05° and the radio contours of the whole SNR are overlaid in white.}
\end{figure*}

Since the best-fit position of the $\gamma$-ray emission is consistent with that of the PWN at high energies, we first assume that the $\gamma$-ray emission comes only from the PWN. We model its spectrum as a power law and we perform a likelihood fit where the spectral parameters of the PWN, those of the nearest point source, the Galactic and isotropic diffuse emissions are free during the fit. The spectral parameters of the other sources are fixed to their best-fit values since they are further than $\sim$ 2° from G326.3$-$1.8. The fit gives a TS value of TS = 593.4, as reported in Table~\ref{tab:LL}, with the number of additional free parameters compared to the model without source. 

 \begin{table}[th!]
	\centering
	\begin{tabular}{l|ccc}
    	\hline
        \hline
        	\rule{0pt}{2.ex} 
			Spatial models 				& TS 		& N$_{\rm{dof}}$ &  TS$_{\rm{PWN}}$ \\
		\hline
        \rule{0pt}{2.ex} 
        	Radio PWN					& 593.4	& 2			& --- \\
            \hline
        \rule{0pt}{2.ex} 
        	Point source 				&  503.3     & 4
        &  --- \\
        \rule{0pt}{2.ex}
        	Point source + radio PWN	& 661.4	& 6
        &  158.1  \\
            \hline
        \rule{0pt}{2.ex} 
		 	Disk 						& 681.8 	& 5 		&  ---  \\
        \rule{0pt}{2.ex}
			Disk + radio PWN 			& 694.8	& 7 		& 13.0 \\
            \hline
        \rule{0pt}{2.ex} 
        	Radio SNR 					& 667.3 	& 2 		& --- \\
        \rule{0pt}{2.ex}
            Radio SNR + radio PWN 		& 683.0	& 4			& 15.7\\
             \hline
        \rule{0pt}{2.ex} 
            SNR mask 					& 670.3	& 2			& ---\\
        \rule{0pt}{2.ex}
            SNR mask + radio PWN 		& 696.4 	& 4 		& 26.1\\
        \hline
	\end{tabular}
    \vspace{0.3cm} 
   	\caption{\label{tab:LL}\small TS values for different spatial models fitted from 1 GeV to 300 GeV. The corresponding number of degrees of freedom is also given (N$_{\rm{dof}}$). TS$_{\rm{PWN}}$ quantifies the improvement of the fit when adding the PWN component to each of the one-component models.}
 \end{table}
Figure~\ref{fig:TSmap_PWN} (left) depicts the 1° $\times$ 1° residual TS map from 1 to 300 GeV obtained by fixing the spectral parameters of the radio PWN and the nearest point source to their best-fit values. The TS map tests a putative point source. It shows qualitatively where there is missing signal, and extended emission can only be more significant than the peak of the TS map. The maximum TS value of the map is TS $\approx$ 60 indicating that this residual emission is clearly significant. The radio template of the PWN is thus not sufficient to describe the data. This confirms our previous results in the Section~\ref{sec:E_dpd_morpho}, that show that the emission below 3 GeV lies outside of the radio contours of the PWN. 

To model the contribution of an additional component that seems to give rise to the low-energy part, we test several templates using first a simple disk component and then physically motivated templates (derived from the radio map of the SNR). We first use the \verb|pointlike| framework to find the best position and extension of an additional source, described by a disk, when the PWN is already included in the model. The fit localizes the position near the center of the SNR at RA$_{\rm{J2000}}$ = $238.169^\circ$ $\pm$ $0.013^\circ$ and Dec$_{\rm{J2000}}$ = $-56.133^\circ$ $\pm$ $0.014^\circ$ with a radius $r$ = $0.295^\circ$ $\pm$ $0.013^\circ$, similar to the radio extension of the SNR ($0.31^\circ$). The significance of the extension is 5.8$\sigma$ (TS$_{\rm ext}$= 33.4), calculated with the TS value of the model including the point source and the radio PWN (reported in Table~\ref{tab:LL}). This rules out the hypothesis of a point-like source being responsible for the additional emission such as an active galactic nucleus behind the SNR. 

We further use the radio observations to derive two other templates for the SNR. First, we use the radio map replacing the contribution of the nebula by the average value of the radio emission around it (labeled ``radio SNR''; see Figure~\ref{fig:Templates}, center). From that, we also create another template, following the radio shock and filled homogeneously (called here ``SNR mask''; see Figure~\ref{fig:Templates}, right).  
 \begin{table*}[th!] 
	\centering
	\begin{tabular}{l|c|cc|cc}
    	\hline
        \hline
        \rule{0pt}{2.ex} 
        Values from the fit  &  Disk only & radio PWN & Disk & radio PWN & SNR mask\\
		\hline
        \rule{0pt}{2.ex} 
        $\Phi$ (ph cm$^{-2}$ s$^{-1}$) $\times$ $10^{-8}$  &  2.70 $\pm$ 0.15
        & 0.23 $\pm$ 0.11 & 2.43 $\pm$ 0.23 
        & 0.33 $\pm$ 0.11 & 2.31 $\pm$ 0.23\\
        \rule{0pt}{2.ex} 
		 $\Gamma$  & 2.07 $\pm$ 0.04  
        &   1.74 $\pm$ 0.15 & 2.16 $\pm$ 0.06 
        &  1.79 $\pm$ 0.12 & 2.17 $\pm$ 0.06 \\
        \hline
	\end{tabular}
   	\caption{\small Results from our maximum likelihood fit between 300 MeV and 300 GeV with the associated statistical errors. $\Phi$ is the integrated flux and $\Gamma$ is the spectral photon index. \label{tab:Fit}}
 \end{table*}
 \begin{figure*}[th!]
  \includegraphics[scale=0.31]{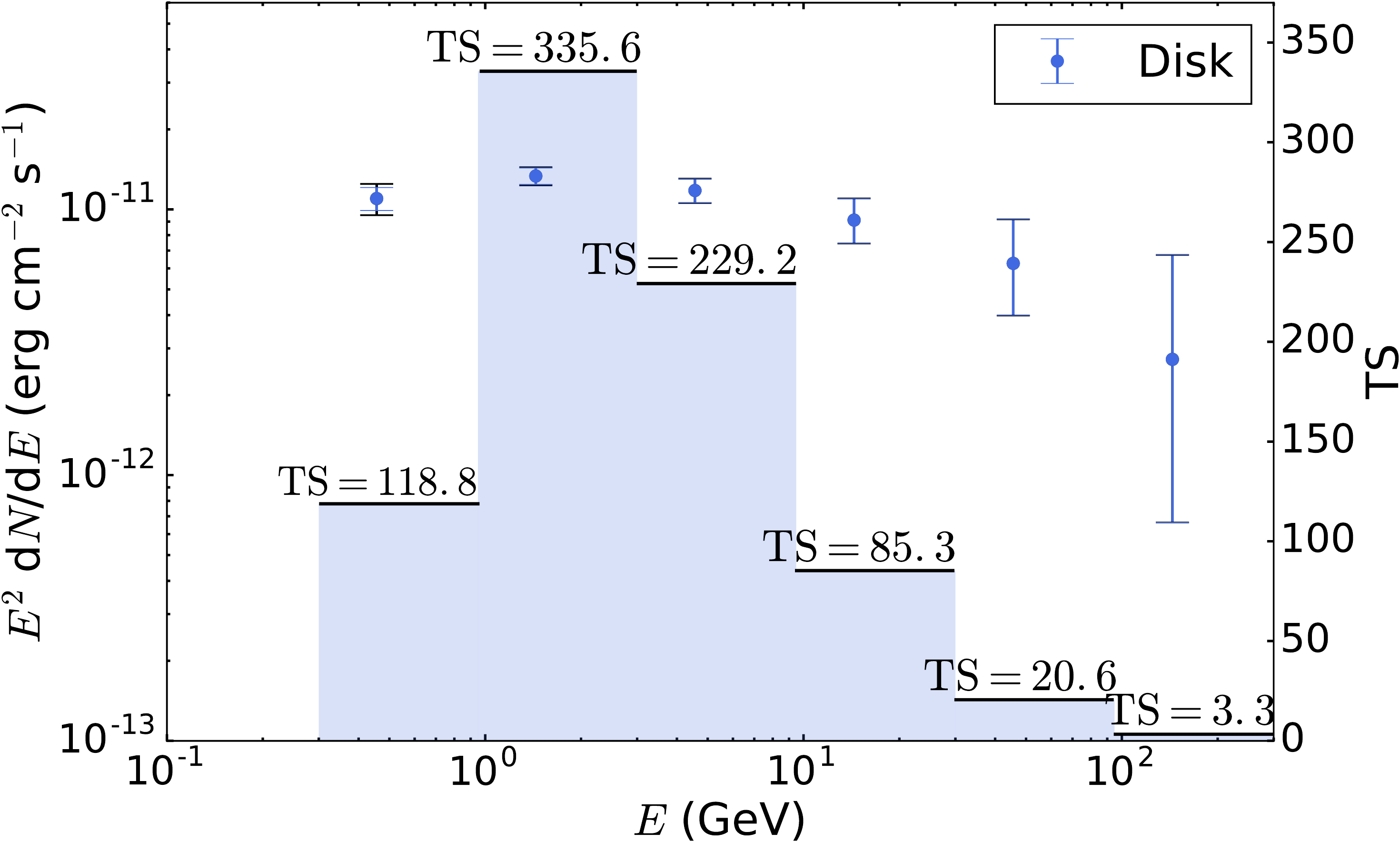}
  \hspace{0.5cm}
    \includegraphics[scale=0.31]{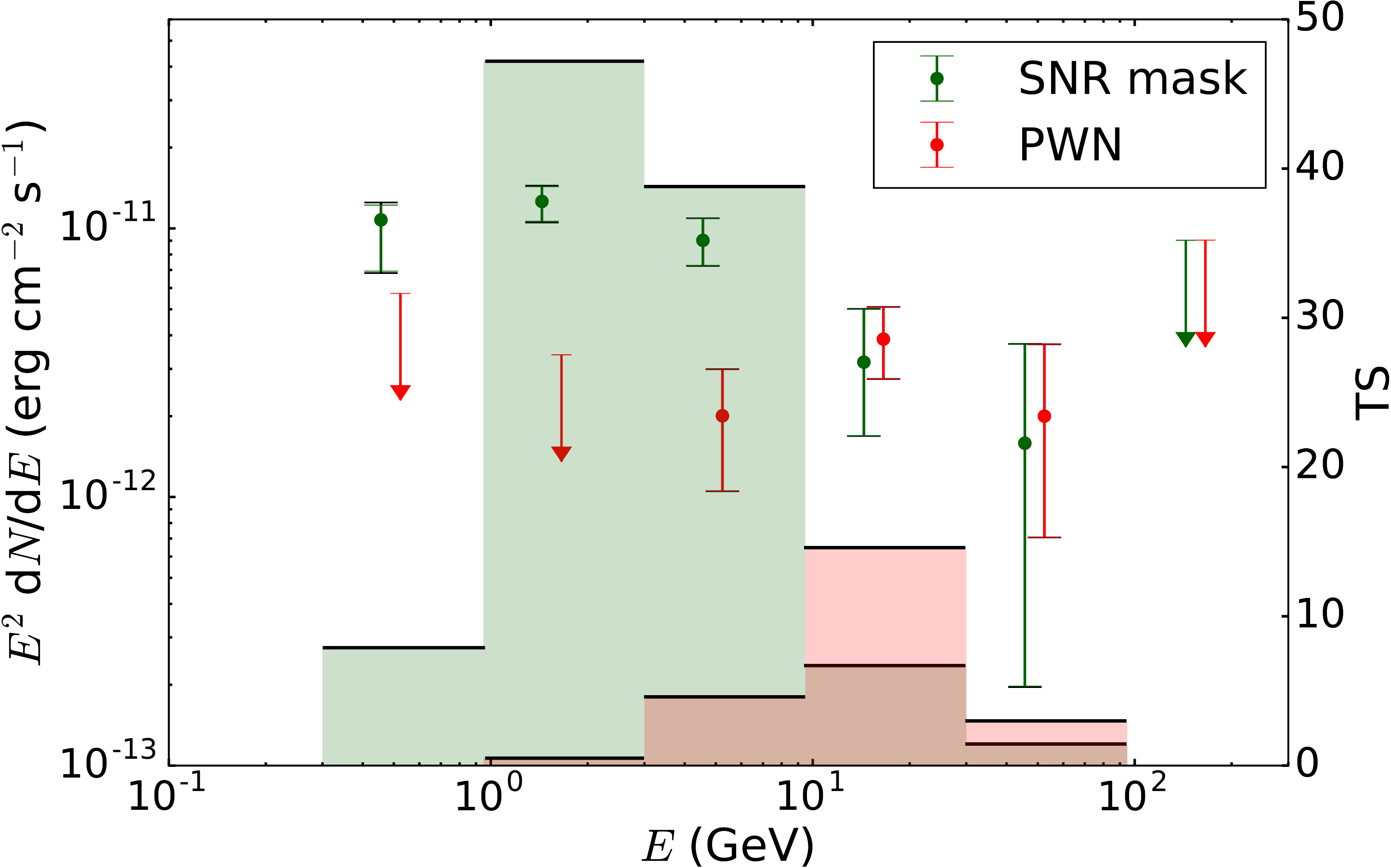}
\caption{\label{fig:SEDs} \small SED (data points) and  TS values (shaded areas) in individual energy bands using the uniform disk model ($\textit{left}$) and using the SNR mask and the radio PWN model ($\textit{right}$). The colored bars are the statistical uncertainties while the black bars correspond to the quadratic sums of statistical and systematic errors (the difference is only visible in the first energy band). The systematic errors are calculated using eight alternative Galactic diffuse emission models.}
\end{figure*}

For all components, the $\gamma$-ray emission is described by a power law and the free spectral parameters are those of the components, the nearest point source, and those of the Galactic and isotropic diffuse emissions. The results from our maximum likelihood fit are given in Table~\ref{tab:LL}  with the numbers of free parameters associated to the models (spectral and/or spatial). \\
First, the TS values obtained using the one-component models (disk, radio SNR or SNR mask) alone are clearly higher than that obtained using only the radio PWN, indicating that the fit prefers a model more extended than the radio PWN. For comparison, Figure~\ref{fig:TSmap_PWN} (right) depicts the residual TS map when only the SNR mask is included to the model, showing a residual emission  coincident with the position of the PWN. In terms of test statistic, when adding a second component, the model with one component becomes the null hypothesis to test the significance of the second component. In Table~\ref{tab:LL}, we compare the TS values obtained with each of the one-component models to the two-component models testing the improvement of the fit when adding the radio template of the PWN. The difference TS$_{\rm{PWN}}$ can be converted to a significance since in the null hypothesis (no PWN emission) it behaves as a $\chi^{2}$-law with two degrees of freedom.

For all our extended models, the significance of adding the PWN lies between 3 and 4$\sigma$ and the maximum TS values are obtained for the model including the radio PWN with either the disk or the SNR mask. In terms of significance, our best model involves the radio PWN and the SNR mask since it requires fewer free parameters during the fit than the disk whose spatial components have been optimized. The lower TS value using the two radio templates (for the SNR and the PWN) indicates that the $\gamma$-ray emission does not entirely follow the synchrotron distribution and the fit prefers a more homogeneous structure for the shell, keeping in mind that this conclusion depends on the model we choose for the PWN.

\subsection{Spectral analysis\label{sec:spectral_analysis}}

To understand the underlying emission processes, we perform a spectral analysis from 300 MeV to 300 GeV using our best models found in the previous section: the disk alone and the radio PWN with either the SNR mask or the disk. Here the $\gamma$-ray emissions are still described with power laws since other spectral representations did not improve the fit. We do not take into account the energy dispersion (this induces a bias $\sim$ $-$ 5\% on flux\footnote{See \url{https://fermi.gsfc.nasa.gov/ssc/data/analysis/documentation/Pass8_edisp_usage.html}}).
Using \verb|gtlike|, we perform a maximum likelihood fit leaving the same spectral parameters free as before. Table~\ref{tab:Fit} reports the results obtained from the fit. 

When we use the disk alone to describe the $\gamma$-ray emission, the photon index is found to be close to 2. Using differentiated models for the PWN and the SNR, the fit leads to a spectral separation between the two components: a softer spectrum for the remnant ($\Gamma \approx 2.16$ using the disk and $\Gamma \approx 2.17$ using the SNR mask) and a harder spectrum for the nebula ($\Gamma \approx 1.74$ and $\Gamma \approx 1.79$). The choice of the model for the remnant (either the disk or the SNR mask) has a very slight impact on the spectral study.

To compute the spectral energy distribution (SED), we divide the whole energy range (300 MeV -- 300 GeV) into six bins and impose a TS threshold of 1 per energy bin for the flux calculation; otherwise an upper limit is calculated. In each bin, the photon indexes of the sources of interest are fixed to 2 to avoid any dependence on the spectral models. The fluxes of the PWN and the SNR components are let free during the fit as well as those of the Galactic and isotropic diffuse emissions. All other sources are fixed to their best global model.

Figure~\ref{fig:SEDs} shows the SED of the uniform disk (left) and the SED of the best-fit two-component model (right) using the radio PWN and the SNR mask. The colored error bars represent the statistical errors while the quadratic sums of the statistical and systematic errors \citep[calculated using eight alternative Galactic diffuse emission models as explained in the first $\textit{Fermi}$-LAT supernova remnant catalog,][]{First_SNRcat:2016} are represented with black horizontal bars. The systematic errors are never dominant and are comparable to the statistical ones only in the first band. Note that the effective area uncertainty also induces systematic errors (10\% between 100 MeV and 100 GeV). 
These SEDs clearly emphasize that two distinct morphologies give rise to two distinct spectral signatures, while the different emissions seem to be mixed when we use a single component model. The TS values in each energy bin highlight the different contributions of the two components: at low energy (E $<$ 10 GeV), the emission is dominated by the SNR while the contribution of the PWN becomes important above 10 GeV, bringing out the spatial and spectral distinctions between these two nested objects.

\section{Results and discussion}
For this entire section we assume the distance to the SNR is 4.1 kpc \citep{Temim:2013}.
\subsection{SNR spectrum}
\afterpage{
\begin{table*}[t!]
	\centering
	\begin{tabular}{l|lcccccccccc}
    	\hline
        \hline
        \rule{0pt}{2.ex}
        Scenario   & Model & $B$ ($\mu$G) & $W_{\rm{p}}$ (erg) & 							$K_{\rm{e-p}}$  & $\Gamma_{\rm{e,1}}$/$\Gamma_{\rm{e,2}}$  & 							$E_{\rm{b}}$ (TeV) & $E_{\rm{max,e}}$ (TeV) &  		$\Gamma_{\rm{p}}$ & $E_{\rm{max,p}}$ (TeV) & 
        				$n_{0}$ (cm$^{-3}$) \\
		\hline
        \rule{0pt}{2.ex} 
     Leptonic & fitted &	10 	& 5 $\times$ $10^{49*}$ &							1.4      &  1.8/2.8      & 
     					(0.15 -- 0.35) & (0.4 -- 1.1) &
                     	2$^{*}$ & ($\geq$ $E_{\rm{max,e}}$) & 
                        0.1$^{*}$ \\
                        & fitted &  20 & 5 $\times$ $10^{49*}$ &
						0.5     &   1.8/2.8        & 
                        (0.4 -- 0.9) & (0.9 -- 1.5) &  
                        2$^{*}$ & ($\geq$ $E_{\rm{max,e}}$)  &
                		0.1$^{*}$ \\
                & 		consistent & 20 & 5 $\times$ $10^{49*}$ &
						0.5       &			1.8/2.8         & 
                        1.9$^{*}$ & 2.3$^{*}$ & 
                 		2$^{*}$ & 2.7$^{*}$ & 
                        0.1$^{*}$ \\
                    \hline
        \rule{0pt}{2.ex} 
    Hadronic  \\
    $\hspace{0.3cm}$ main shock  		& consistent & 10 & 5 $\times$ $10^{49}$ & 
    					0.03   &  2/3$^{*}$ & 
                    	1.4$^{*}$ &  1.4$^{*}$ &  
                    	2$^{*}$ &  1.4$^{*}$ & 
                        0.1$^{*}$  
         \\
 $\hspace{0.3cm}$ radiative shock	& consistent & 13.6$^{*}$ &  & 
                       &   & 
                      &  & 
                      &  &
                     1.88$^{*}$
		 \\
        \hdashline
        \rule{0pt}{2.ex}
 $\hspace{0.3cm}$ cooled regions & & 158$^{*}$ &  1.9 $\times$ $10^{49}$   &
						 0.03 &   1.8/2.8     &
                         0.02$^{*}$ & 0.08$^{*}$  &
                          2$^{*}$   &   0.08$^{*}$        &
                             88.3$^{*}$ \\
        \hline
	\end{tabular}
   	\caption{\small Required physical parameters to model the radio and the $\gamma$-ray data coming from the shell in case of leptonic and hadronic scenarios. The parentheses indicate the range of permitted values while the asterisks indicate the fixed values in our analysis. The other values (without parentheses and asterisks) are adjusted by hand. In the two leptonic scenarios in which the break energy $E_{\rm{b}}$ and maximum energy $E_{\rm{max,e}}$ of the electrons are fitted, they are not consistent with the magnetic field if they are due to synchrotron cooling. In all other (consistent) models, they are calculated following \cite{Parizot:2006}. For the hadronic scenario, the first and the second line correspond to the properties of the main shock ($u_{\rm{sh}}$ = 500 km s$^{-1}$) and the radiative shock ($u_{\rm{sh,cl}}$ = 150 km s$^{-1}$), respectively. The density and the magnetic field in the last line are those of the downstream cooled regions. \label{tab:Phys_param}}
 \end{table*}
 
\begin{figure*}[t!]
\includegraphics[width=0.5\textwidth]{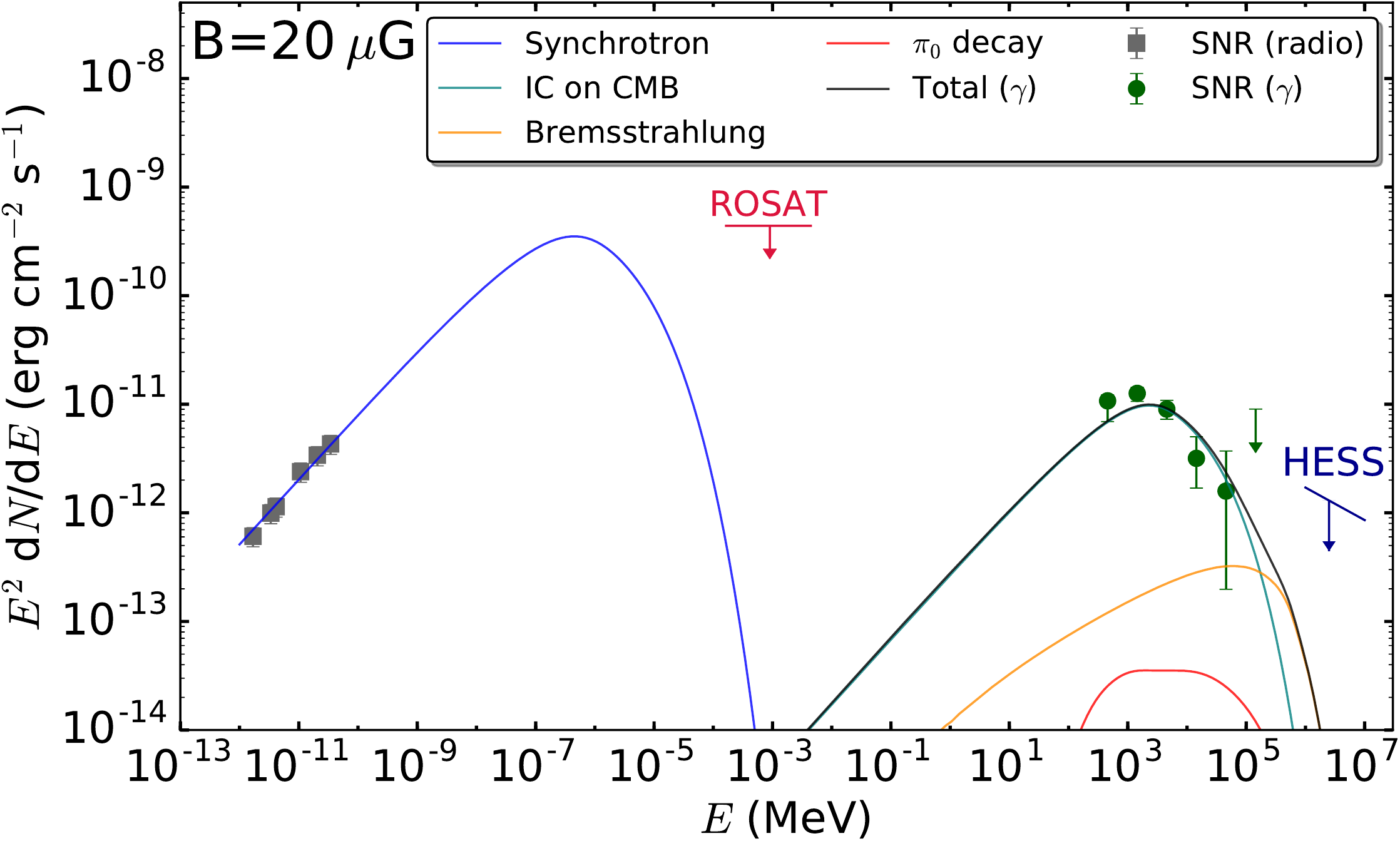}
\includegraphics[width=0.5\textwidth]{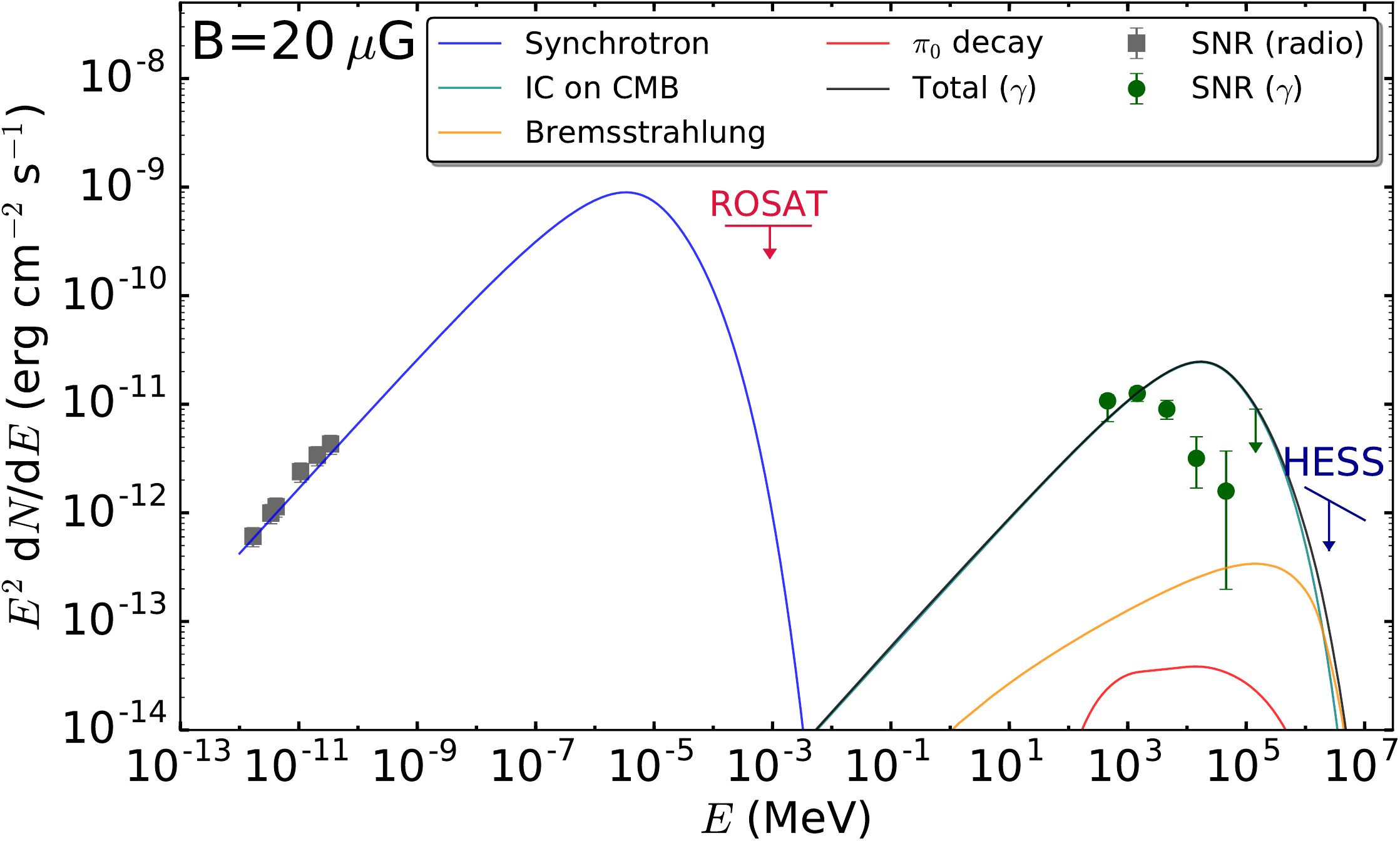}
\caption{\label{fig:Fit_naima_SNR_Lepto} \small Multi-wavelength modeling of the SNR spectrum in the leptonic scenario. The radio points are extracted from \cite{Dickel:2000}, the \textit{ROSAT} and the H.E.S.S upper limits come from \cite{Kassim:1993} and \cite{HESS:SNRpop:2018}, respectively. (\textit{Left}) Parameters are free to vary and values are: $E_{\rm{b}}$ = 600 GeV, $E_{\rm{max,e}}$ = 1 TeV, $E_{\rm{max,p}}$ = 1 TeV, $W_{\rm{p}}$ = 5 $\times$ 10$^{49}$ erg and $K_{\rm{e-p}}$ = 0.5. The spectral index of the electrons before and after the break is $\Gamma_{\rm{e,1}}$ = 1.8 and $\Gamma_{\rm{e,2}}$ = 2.8, respectively. (\textit{Right}) Same as in the left panel but with values consistent with the magnetic field: $E_{\rm{b}}$ = 1.9 TeV, $E_{\rm{max,e}}$ = 2.3 TeV and $E_{\rm{max,p}}$ = 2.7 TeV.}
\end{figure*}
\begin{figure*}[th!]
\hspace{0.7cm}
\includegraphics[scale=0.37]{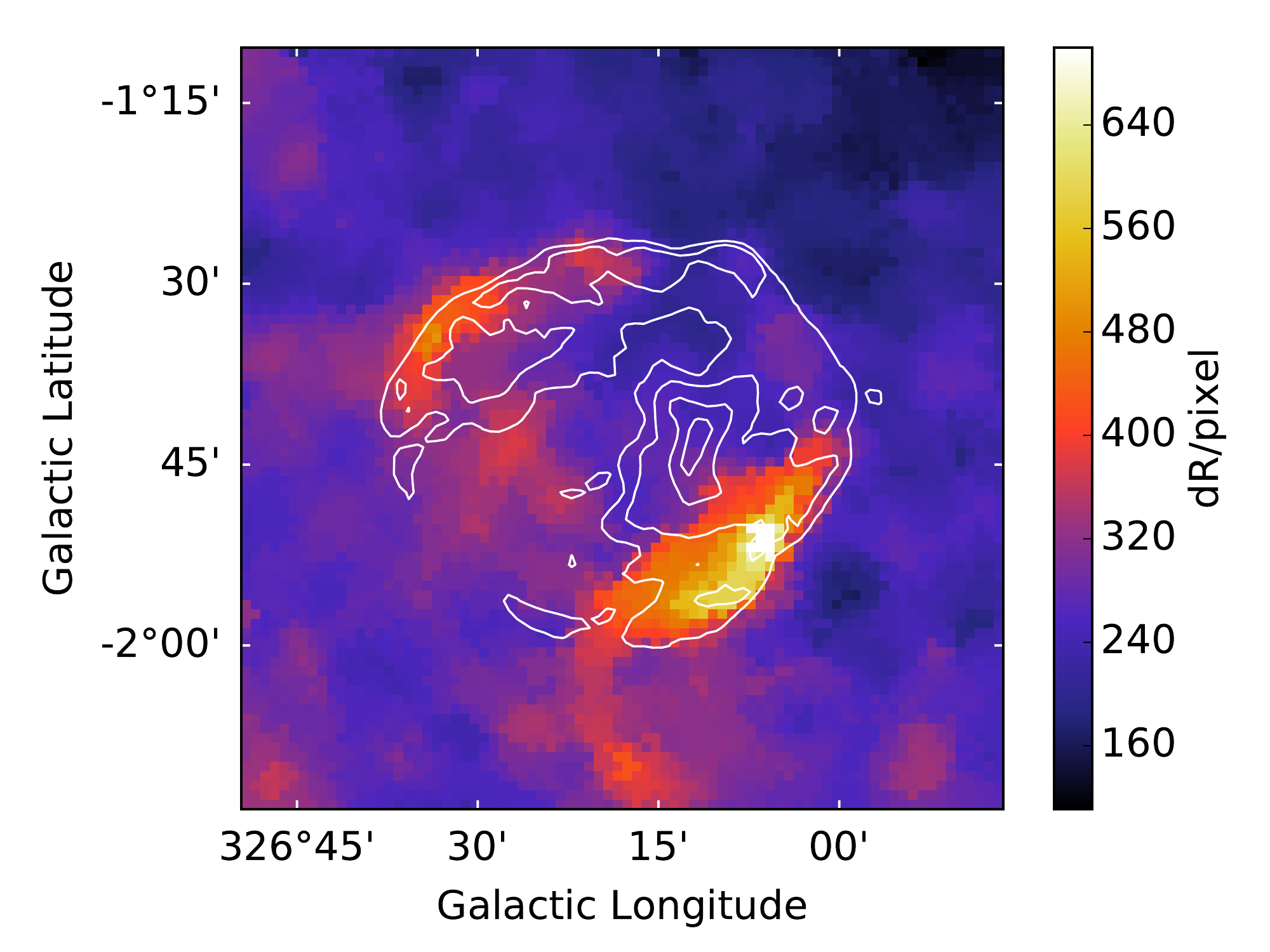}
\hspace{0.7cm}
\includegraphics[width=0.5\textwidth]{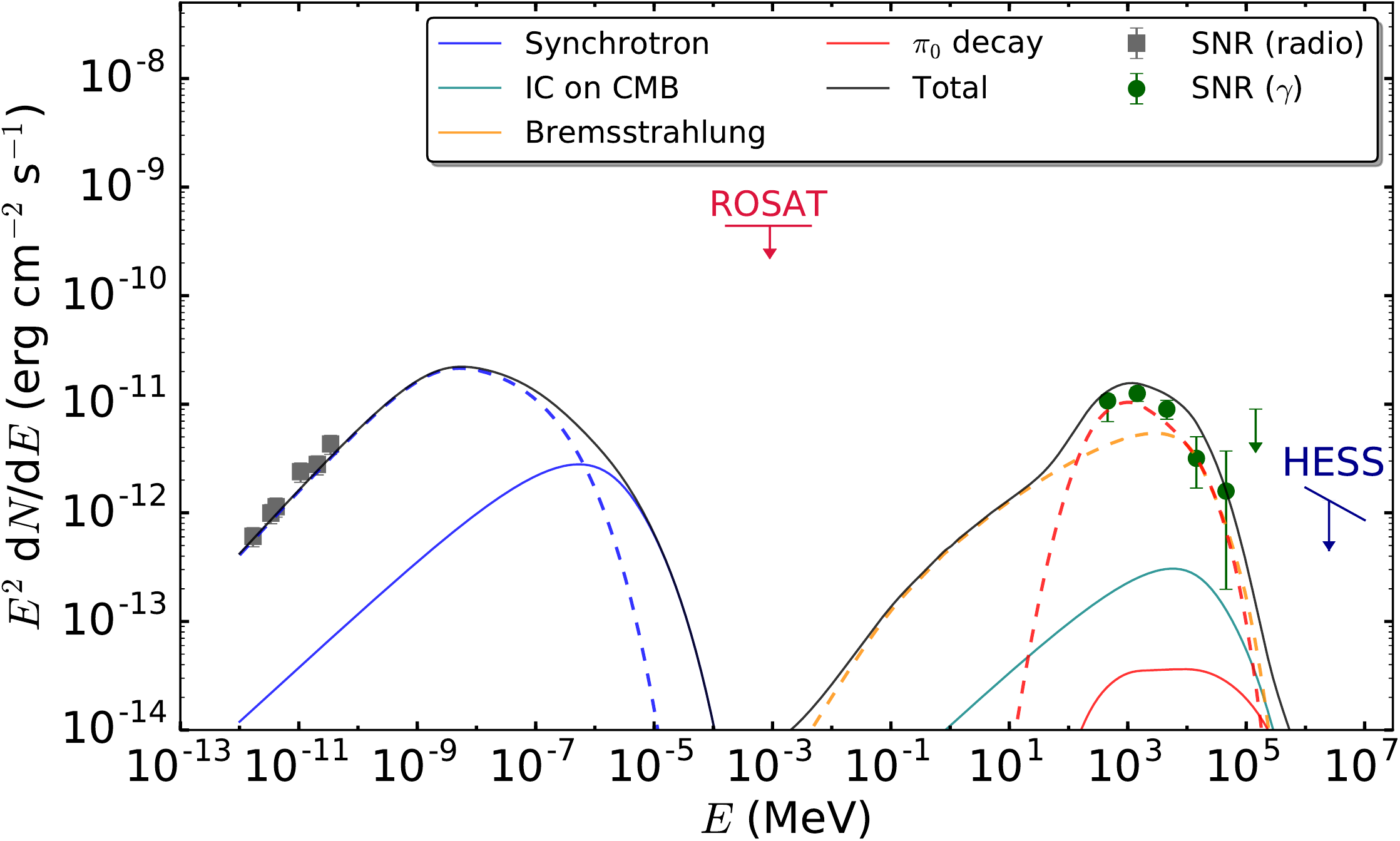}
\caption{(\textit{Left}) \small H$\alpha$ emission of SNR G326.3$-$1.8 obtained from the Southern H-Alpha Sky Survey Atlas \citep{Gaustad:2001}. The radio contours of the whole SNR are overlaid in white. (\textit{Right}) Multi-wavelength modeling in the hadronic scenario with two contributions: a main shock (solid lines) and a radiative shock (dashed lines). The radio points are extracted from \cite{Dickel:2000}, the \textit{ROSAT} and the H.E.S.S upper limits come from \cite{Kassim:1993} and \cite{HESS:SNRpop:2018}, respectively. The values of the parameters are reported in Table~\ref{tab:Phys_param}. \label{fig:Naima_Hadro_Halpha}}
\end{figure*}
}

To understand the observed $\gamma$-ray spectrum of the SNR, we perform multi-wavelength modeling using the one-zone models provided by the \verb|naima| package \citep{naima}.  
From \cite{Dickel:2000} we take the five radio flux measurements of the shell. As there is no associated synchrotron emission in the X-ray domain, we use the \textit{ROSAT} thermal flux reported by \cite{Kassim:1993} as an upper limit. We also use the TeV upper limit derived from H.E.S.S. with 14 hours of observational live time and assuming a photon index of 2.3 \citep{HESS:SNRpop:2018}.

Assuming the Sedov phase, we derive the kinetic energy released by the supernova:
\begin{equation}
\hspace{0.5cm}
\frac{E_{\rm{SN}}}{10^{51} \hspace{0.1cm} \rm{erg}} = R_{12.5}^5 \times \Big(\frac{n_{0}}{ \textrm{cm}^{-3}}\Big) \times t_{4}^{-2} = 0.5
\end{equation}
  where $R_{12.5}$ = $R$/(12.5 pc) and $t_{4}$ = $t$/(10,000 yrs) taking $R$ = 21 pc, $t$ = 16,500 yrs and $n_{0}$ = 0.1 cm$^{-3}$ for a distance of 4.1 kpc. We take the inputs from \cite{Temim:2013} but do not derive the same explosion energy. Here we use the common values of $\xi$ = 2.026 (for $\gamma$ = 5/3) and $\rho_{0}$ = 1.4$m_{\rm{H}}n_{0}$ in the usual Sedov equation: $R^{5}$ = $\xi$ ($E_{\rm{SN}}$/$\rho_{0}$) $t^{2}$. 
  
The explosion energy and age depend on the distance, which is still uncertain. In addition, the density and temperature (the latter provides the shock speed estimate) were derived from a small region in the south of the SNR and it is not yet clear whether this region is representative of the rest of the SNR. (A Large Program with \textit{XMM-Newton} on G326.3$-$1.8 is currently ongoing and will provide more constraints on the thermal emission across the SNR.) In this regard, the multi-wavelength modeling presented in this section is not to be viewed as a precise measurement of the properties of the accelerated particles but rather as showing that a simple self-consistent model can reproduce the observations.

We describe the electron population as a broken power law spectrum with spectral indexes $\Gamma_{\rm{e,1}}$/$\Gamma_{\rm{e,2}}$ with an exponential cut-off. The break at energy $E_{\rm{b}}$ is assumed to be due to cooling so we set $\Gamma_{\rm{e,2}}$ = $\Gamma_{\rm{e,1}}$ + 1 while the cut-off defines the maximum attainable energy of the particles $E_{\rm{max,e}}$. The proton spectrum is described as a power law with spectral index $\Gamma_{\rm{p}}$ with an exponential cut-off $E_{\rm{max,p}}$. In our models we consider by default the CMB as the only photon seed for IC scattering. \\
\subsubsection{Leptonic scenario}
We first investigate the leptonic scenario for which we vary the values of the magnetic field $B$, the total energy budget in electrons $W_{\rm{e}}$ and protons $W_{\rm{p}}$, and the break and maximum energy of the particles. Figure~\ref{fig:Fit_naima_SNR_Lepto} (left) shows one of the combinations that simultaneously fits the radio and the $\gamma$-ray data. Since this solution is not unique, we report in Table~\ref{tab:Phys_param} the range of permitted values of these parameters. For clarity, we fix the total energy in protons to W$_{\rm{p}}$ = 5 $\times$ 10$^{49}$ erg (corresponding to 10\% of $E_{\rm{SN}}$) and we report the range of permitted values of the electron-proton ratio $K_{\rm{e-p}}$. Since the maximum energy of protons is always higher than that of electrons, which suffer synchrotron losses, we use the maximum value of $E_{\rm{max,e}}$ as a lower limit for $E_{\rm{max,p}}$. 

To reproduce the radio spectral shape, we need a hard index for the electrons $\Gamma_{\rm{e,1}}$ = 1.8 (taking thus $\Gamma_{\rm{e,2}}$ = 2.8), while we keep $\Gamma_{\rm{p}}$ = 2 due to the lack of observational constraints.
For a $B$ field between 10 and 20 $\mu$G, the $\gamma$-ray data can only be explained if the total energy in electrons reaches $W_{\rm{e}}$ = (2.5 -- 7) $\times$ $10^{49}$ erg, which is clearly unreasonable since that requires a $K_{\rm{e-p}}$ between 0.5 and 1.4. If in order to reduce $W_{\rm{e}}$, we increase $B$ to higher values than expected for the compressed ISM, the $\gamma$-ray data cannot be fitted and the IC spectrum lies one order of magnitude below the data. Even if infrared and optical photon fields with energy density 0.26 eV cm$^{-3}$ each (the same as the CMB value, which is a reasonable estimate 100 pc below the Galactic plane) are added, an unrealistically large $W_{\rm{e}}$ is still required to fit the $\gamma$-ray data.

Another inconsistency of that model is that the values of $E_{\rm{b}}$, $E_{\rm{max,e}}$ and $E_{\rm{max,p}}$ reported in Table~\ref{tab:Phys_param} are not consistent with the magnetic field. Following \citet{Parizot:2006}, we use the synchrotron loss time:
\begin{equation}
\centering
\hspace{0.5cm} \tau_{\rm{sync}}=(1.25 \times 10^{3}) \times E_{\rm{TeV}}^{-1} B_{100}^{-2} \hspace{0.2cm} \textrm{yrs}
\end{equation}
and the acceleration timescale:
\begin{equation}
\hspace{0.5cm} t_{\rm{acc}}= 30.6 \times \frac{3r^2}{16(r-1)} \times k_{\rm{0}}(E) \times E_{\rm{TeV}}B_{100}^{-1}u_{\rm{sh,3}}^{-2} \hspace{0.2cm} \textrm{yrs}
\end{equation}
with $r$ being the shock compression ratio, $k_0$ the ratio between the mean free path and the gyroradius, $B_{\rm{100}}$ and $u_{\rm{sh,3}}$ the magnetic field and the shock velocity in units of 100 $\mu$G and 1000 km s$^{-1}$, respectively. $k_{\rm{0}}$ $\geq$ 1 can be interpreted as the ratio of the total magnetic energy density to that in the turbulent field ($B_{\rm{tot}}^{2}/B_{\rm{turb}}^{2}$) and $k_0$ $\approx$ 1 has been found for young SNRs \citep{Uchiyama:2007}. For evolved systems, we expect the turbulent magnetic field to be smaller than the large-scale component (so that $k_0$ $>$ 1) and we adopt here $k_0$ = 10 for the highest-energy electrons. Taking $r$ = 4, we thus calculate $E_{\rm{b}}$ (equating $\tau_{\rm{sync}}$ = $t_{\rm{age}}$), $E_{\rm{max,e}}$ ($t_{\rm{acc}}$ = min$\{\tau_{\rm{sync}}, t_{\rm{age}}\}$) and $E_{\rm{max,p}}$ ($t_{\rm{acc}}$ = $t_{\rm{age}}$). For $B$ = 20 $\mu$G, we obtain $E_{\rm{b}}$ = 1.9 TeV, $E_{\rm{max,e}}$ = 2.3 TeV  and $E_{\rm{max,p}}$ = 2.7 TeV. Figure~\ref{fig:Fit_naima_SNR_Lepto} (right) shows the corresponding spectrum which implies an IC cut-off at too high energy and does not fit the $\gamma$-ray data. 

One noteworthy aspect of this source concerns the difficulty to explain its very high radio flux \citep[114 Jy at 1 GHz,][]{Dickel:2000}. High radio fluxes are also found in middle-aged SNRs interacting with molecular clouds, such as W44 \citep[230 Jy at 1 GHz,][]{Castelletti07:W44} and IC 443 \citep[160 Jy at 1 GHz,][]{Milne71:IC443}, where the highly compressed gas enhances the synchrotron emission. 

The particularly high total energy required in electrons to reproduce the SNR spectrum and the impossibility to fit the data with consistent values rule out a leptonic origin of the $\gamma$-ray emission and lead us to investigate the hadronic scenario. 

\subsubsection{Hadronic scenario}

\cite{Vandenbergh:1979} has reported H$\alpha$ emission in the northeast and southwest regions of the SNR (see Figure~\ref{fig:Naima_Hadro_Halpha}, left panel) and \cite{Dennefeld:1980} obtained a spectrum indicating an [S II]/H$\alpha$ ratio characteristic of a radiative shock. This is evidence of the interaction of the shock with neutral material where some regions of the SNR are entering the radiative phase while other parts are freely expanding in the ISM. As a consequence, we suggest to model the SNR spectrum with two contributions:
\begin{itemize}
 \item a radiative shock arising from the presence of clouds in the surroundings of the SNR
 \item a main shock with a velocity of $u_{\rm{sh}}$ = 500 km s$^{-1}$ and expanding in an ISM density of $n_{\rm{0}}$ = 0.1 cm$^{-3}$ \citep{Temim:2013}
\end{itemize}

Below we calculate the physical parameters associated with the radiative component. \cite{Uchiyama:2010} studied the non-thermal emission from crushed clouds in SNRs where re-acceleration of pre-existing cosmic rays can explain the observed GeV emission powered by hadronic interactions. \\
Following this work, the strong shock driven into the clouds has a velocity of: 
\begin{equation}
\hspace{0.5cm}
u_{\rm{sh,cl}} = k \sqrt{\frac{n_0}{n_{\rm{0,cl}}}} \times u_{\rm{sh}}
\end{equation}
where $k$ = 1.3 is adopted as in \cite{Uchiyama:2010}, $n_{0}$ and $n_{\rm{0,cl}}$ being the upstream ISM and clouds density respectively. 
For the upstream magnetic field in the clouds, we have:
\begin{equation}
\hspace{0.5cm}
B_{\rm{0,cl}} = b \sqrt{\frac{n_{\rm{0,cl}}}{\textrm{cm}^{-3}}} \hspace{0.1cm} \mu \textrm{G} \label{Bup}
\end{equation}
where $b$ = $v_{\rm{A}}$/(1.84 km s$^{-1}$) with $v_{\rm{A}}$ being the Alfvén velocity, the mean value of which is thought to be roughly equal to the velocity dispersion observed in molecular clouds ($\sim$ 0.5 -- 5 km s$^{-1}$), implying $b$ $\sim$ 0.3 -- 3 \citep{HM:89}. 
As in \cite{Uchiyama:2010}, we assume that the magnetic pressure in the cooled gas is equal to the shock ram pressure and we have:
\begin{equation}
\hspace{0.5cm}
\frac{B_{\rm{m}}^2}{8 \pi} = k^2  n_{\rm{0}}  \mu_{\rm{H}} u_{\rm{sh}}^2 \label{press_eq}
\end{equation}
where $\mu_{\rm{H}}$ is the mass per hydrogen nucleus and $B_{\rm{m}}$ the downstream magnetic field in the cooled regions:
\begin{equation}
\hspace{0.5cm}
B_{\rm{m}} = \sqrt{\frac{2}{3}} \times \Big(\frac{n_{\rm{m}}}{n_{\rm{0,cl}}}\Big) \times B_{\rm{0,cl}} \label{Bcool}
\end{equation}
with $n_{\rm{m}}$ being the downstream density in the cooled regions.
In this model, the compressed magnetic field is fixed to 158 $\mu$G by the pressure in the SNR (Eq~\ref{press_eq}). This requires a large $W_{\rm{e}}$ in the clouds for the synchrotron emission to be consistent with the bright observed radio flux. We set $K_{\rm{e-p}}$ in the clouds to 0.03, which is large but still reasonable, since a lower $K_{\rm{e-p}}$ would result in an uncomfortably large W$_{\rm{p}}$. In any case, the cosmic-ray energy in the shocked clouds must be high. Thus, fitting simultaneously the $\gamma$-ray data and the radio data implies that the compressed density should be relatively low. From Eqs (\ref{Bup}, ~\ref{press_eq}, ~\ref{Bcool}), we have: 
\begin{equation}
\hspace{0.5cm}
n_{\rm{m}} = \sqrt{\frac{3}{\pi \mu_{\rm{H}}}} \times \frac{B_{\rm{m}}^2}{4 b \times u_{\rm{sh,cl}}}
\end{equation}
We adopt here the highest reasonable values $b$ = 3 and $u_{\rm{sh,cl}}$ = 150 km s$^{-1}$ (above which the shock would have no time to become radiative). Thus the downstream density in the cooled regions is $n_{\rm{m}}$ = 88.3 cm$^{-3}$.
Taking $n_{\rm{0}}$ = 0.1 cm$^{-3}$, $u_{\rm{sh}}$ = 500 km s$^{-1}$ \citep{Temim:2013}, $u_{\rm{sh,cl}}$ = 150 km s$^{-1}$ and $b$ = 3, we obtain for the upstream density and magnetic field in the clouds $n_{\rm{0,cl}}$ = 1.88 cm$^{-3}$ and $B_{\rm{0,cl}}$ = 4.11 $\mu$G. This relatively low density is in agreement with the non-detection of CO lines close to this SNR. The densities encountered in G326.3$-$1.8 (cloud and intercloud medium) would then be very similar to the Cygnus Loop \citep{Raymond:1988}.

The electrons accelerated in the clouds will rapidly cool due to the strong magnetic field in the dense regions for which we derive the break energy of the particles by equating $\tau_{\rm{sync}}$ = $t_{\rm{age}}$/2 (time since the clouds were shocked). At the shock front, the downstream magnetic field is $B_{\rm{d,cl}}$ = $\sqrt{11}B_{\rm{0,cl}}$ = 13.6 $\mu$G, assuming a randomly directed field, and we derive the corresponding maximum energy of the particles, using $k_{\rm{0}}$ = 10 and $u_{\rm{sh,cl}}$ = 150 km s$^{-1}$ when equating $t_{\rm{acc}}$ = min$\{\tau_{\rm{sync}},t_{\rm{age}}/2\}$. For particles trapped in the clouds, we thus find $E_{\rm{b}}$ = 15.2 GeV and $E_{\rm{max,e}}$ = $E_{\rm{max,p}}$ = 82.7 GeV. 

Figure~\ref{fig:Naima_Hadro_Halpha} (right) shows the corresponding spectrum with the contributions from the main shock (solid lines) and the radiative shock (dashed lines). With such high magnetic field and density in the cooled regions, the radiative shock dominates the synchrotron and the $\gamma$-ray emission. Setting  $K_{\rm{e-p}}$ = 0.03, the observed spectrum can be explained with $W_{\rm{p}}$ = 1.9 $\times$ $10^{49}$ erg (and thus $W_{\rm{e}}$ = 5.7 $\times$ $10^{47}$ erg), corresponding to 3.8\% of $E_{\rm{SN}}$ transmitted to the re-accelerated protons in the clouds. To reproduce the radio spectral shape, we use harder indexes for the electrons at the radiative shock $\Gamma_{\rm{e,1}}$/$\Gamma_{\rm{e,2}}$ = 1.8/2.8 which is also observed in other radiative SNRs \citep{Ferrand:SNRcat}\footnote{Radio spectral indexes of some radiative SNRs, such as W 44 or IC 443, can be found at \url{http://www.physics.umanitoba.ca/snr/SNRcat/}}.
The $\gamma$-ray cut-off implied by $E_{\rm{max,p}}$ = 82.7 GeV fits the observed spectrum well. This is however largely coincidental. $k_{\rm{0}}$ is unconstrained, $t_{\rm{acc}}$ is unknown (we do not know when the clouds were shocked). \cite{Uchiyama:2010} predict an increase of the maximum energy by a factor of (($n_{\rm{m}}$/$n_{\rm{0}}$)/4)$^{1/3}$ = 2.27 due to adiabatic compression, which we did not enter into $E_{\rm{max,p}}$. The damping of Alfvén waves due to ion-neutral collisions also implies a break in the proton spectrum that we did not take into account because, with $B_{\rm{0,cl}}$ = 4.11 $\mu$G and $n_{\rm{0,cl}}$ = 1.88 cm$^{-3}$, it occurs around 100 GeV. Observationally, $E_{\rm{max,p}}$ must range between 30 and 100 GeV, which is in between other radiative SNRs such as W 44 or IC 443 \citep[respectively 22 and 239 GeV,][]{Ackermann:13}.

Since our model predicts that radiative shocks can explain the entire spectrum, we cannot assess observational constraints at the main shock. We take $B$ = 10 $\mu$G, implying $B_{\rm{ISM}}$ $\approx$ 3 $\mu$G (with $r$ = 4), to stay consistent with $B_{\rm{0,cl}}$ $>$ $B_{\rm{ISM}}$ but $B_{\rm{ISM}}$ could have been lower. We also use the typical 10\% of $E_{\rm{SN}}$ going into protons but this and the value of $K_{\rm{e-p}}$ could also be reduced. For the particle spectra, we keep $\Gamma_{\rm{e,1}}$ = $\Gamma_{\rm{p}}$ = 2 and $\Gamma_{\rm{e,2}}$ = 3 since we have simple acceleration at the main shock and no observational constraints. The corresponding break and maximum energy are calculated following \cite{Parizot:2006} with $u_{\rm{sh}}$ = 500 km s$^{-1}$ and $k_{\rm{0}}$ = 10 as we did for the leptonic-dominated scenario. All the values used for the plot are reported in Table~\ref{tab:Phys_param}.

The entire SNR spectrum can thus be explained by the emission from radiative shocks. Although there is no clear correlation between the H$\alpha$ and the radio maps, this difference can be explained by the orientation of the magnetic field: where $B$ is perpendicular to the shock velocity, the synchrotron emission is largest (compression of the tangential component of the field) whereas optical emission should be enhanced when $B$ is parallel since the compression is no longer limited by the magnetic field. Quantitatively, the total energy required in the cosmic rays at the radiative shocks is large. Assuming 20\% of the pressure in the radiative shocks is in the form of cosmic rays (the rest is mostly magnetic), it requires a surface covering factor close to 50\% (consistent with the fact that we see little deviation from a uniform disk). This may be tested by deep H$\alpha$ imaging. \\

\subsection{PWN spectrum}
\begin{figure}[th!]
\centering
\includegraphics[scale=0.37]{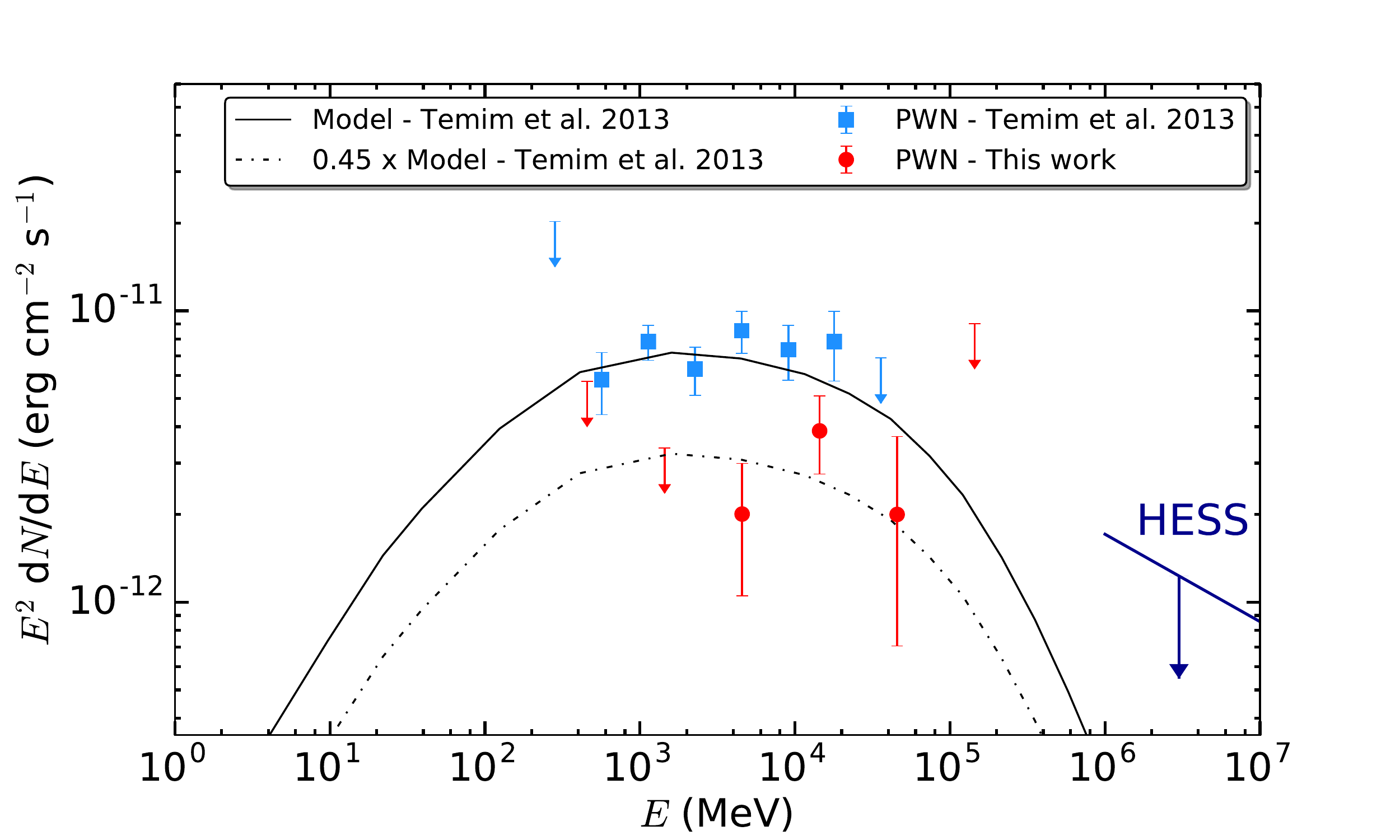}
\caption{\label{fig:PWN_Temim_comparison} \small Comparison of the $\gamma$-ray PWN spectra where the model derived in \cite{Temim:2013} is multiplied by a factor of 0.45 to fit our data.}
\end{figure}
We find that the largest fraction of the $\gamma$-ray emission comes from the SNR, presumably from the hadronic process. Nevertheless the PWN appears to contribute as well. We briefly and qualitatively discuss the impact of the PWN flux diminution on the physical parameters derived in \cite{Temim:2013} who assumed the entire $\gamma$-ray emission originates in the PWN, and based their analysis on the previous data release (Pass 7). 
Figure~\ref{fig:PWN_Temim_comparison} compares the two $\gamma$-ray spectra where the model of \cite{Temim:2013}, who assumed a fully leptonic origin of the emission, is scaled to fit our data. The current flux corresponds to 45\% of the previous one. \\
If we approximate $W_{\rm{e}}$ $\approx$ $\int_{0}^{t_{\rm{age}}} \dot{E} dt$, $\dot{E}$ being the energy loss rate of the pulsar, we obtain: 
\begin{equation}
\hspace{3cm} W_{\rm{e}} \approx \dot{E_{0}}\frac{\tau_{\rm{0}} t_{\rm{age}}}{\tau_{\rm{0}} + t_{\rm{age}}}
\end{equation}
where $\tau_0$ is the initial spin-down timescale of the pulsar and $\dot{E_{0}}$ the initial spin-down power.
\cite{Temim:2013} derived $\tau_{0} \approx 2.1 \times 10^{4}$ years and $\dot{E_{0}}$ = 3 $\times$ 10$^{38}$ erg s$^{-1}$. We now require less than half of $W_{\rm{e}}$ leading to $\dot{E_{0}}$ = 1.35 $\times$ 10$^{38}$ erg s$^{-1}$ for the same age and initial spin-down time scale of the pulsar. \\
In their 1-D model, the observed SNR radius is reached at an age of 19 kyrs for which they estimate the PWN magnetic field to be $B_{\rm{PWN}}$ = 34 $\mu$G. The decrease in $W_{\rm{e}}$ would thus also imply a higher magnetic field to still stay consistent with the radio flux of the PWN. 

However, a more nuanced interpretation is required given the complexity of this object. This will require more investigations and detailed modeling which are beyond the scope of this paper. Note also that the PWN spectrum derived in this analysis is model-dependent when considering the assumption made for the SNR. In any case, its flux is reduced compared to previous studies since the SNR contributes most of the $\gamma$-ray emission.


\section{Conclusions}

We perform an analysis from 300 MeV to 300 GeV of the composite SNR G326.3$-$1.8 with the $\textit{Fermi}$-LAT Pass 8 data. We take advantage of the new PSF3 event class by selecting the events with the best angular reconstruction to limit mixture between the SNR and the PWN contributions and also emission from the Galactic plane.  Using the \verb|pointlike| and the \verb|gtlike| frameworks, we confirm that the emission is significantly extended (more than 13$\sigma$) between 300 MeV and 300 GeV.  We perform an analysis in five energy bands which shows that the morphology evolves with energy and the size shrinks towards the radio PWN at high energies (E $>$ 3 GeV). We thus investigate a more detailed morphology using the radio map of the PWN as a starting point. We find that it is clearly not sufficient to describe the $\gamma$-ray data and that an additional extended component is needed. We then test different models for an additional contribution such as a uniform disk, the radio map of the remnant and its homogeneously filled radio template, called here the SNR mask.  Using the maximum likelihood fitting procedure starting at 1 GeV, we find that the model with the SNR mask and the radio PWN reproduces the $\gamma$-ray emission best.

Modeling both $\gamma$-ray emissions by a power law from 300 MeV to 300 GeV, we obtain a spectral separation between the two components: a softer spectrum for the remnant ($\Gamma=2.17$ $\pm$ $0.06$) and a harder spectrum for the nebula ($\Gamma=1.79$ $\pm$ $0.12$). The corresponding SEDs also highlight their different contributions: the SNR dominates the low-energy part (300 MeV -- 10 GeV) while the PWN protrudes at higher energies (E $>$ 10 GeV). 

Concerning the PWN spectrum, we briefly discuss the impact of the flux diminution (about 55\%) compared to previous studies that assumed that the entire $\gamma$-ray emission may come from the PWN. 

The spectral modeling of the SNR emission disproves the leptonic scenario since it requires an unrealistic high energy budget in the electrons to fit the $\gamma$-ray data ($W_{\rm{e}}$ of several $10^{49}$ erg). As H$\alpha$ emission has been reported in this SNR, we suggest a spectral modeling where the main contribution arises from regions entering the radiative phase. The high magnetic field and density in the cooled regions lead to enhanced synchrotron and GeV emission that dominates the entire spectrum. 
The best-fit model involves a compressed magnetic field of 10 $\mu$G and 158 $\mu$G at the main and the radiative shock respectively. With 3.8\% of the kinetic energy released by the supernova going into particles at the radiative shock, we find that an electron-proton ratio of $K_{\rm{e-p}}$ = 0.03 can adequately reproduce the observed spectrum. Although this ratio is slightly higher than one would expect, this is the most appropriate and consistent model we find that can simultaneously explain the high radio and $\gamma$-ray emissions from this SNR.
In the future, CTA (Cherenkov Telescope Array) will give more insight into the properties of this source, providing better sensitivity above 30 GeV. 

\small{\section*{Acknowledgements}
We thank D. Castro and the referee P. Slane for their helpful comments on this paper. \\
The \textit{Fermi} LAT Collaboration acknowledges generous ongoing support from a number of agencies and institutes that have supported both the development and the operation of the LAT as well as scientific data analysis. These include the National Aeronautics and Space Administration and the Department of Energy in the United States, the Commissariat \`a l'Energie Atomique and the Centre National de la Recherche Scientifique / Institut National de Physique Nucl\'eaire et de Physique des Particules in France, the Agenzia Spaziale Italiana and the Istituto Nazionale di Fisica Nucleare in Italy, the Ministry of Education, Culture, Sports, Science and Technology (MEXT), High Energy Accelerator Research Organization (KEK) and Japan Aerospace Exploration Agency (JAXA) in Japan, and the K.~A.~Wallenberg Foundation, the Swedish Research Council and the Swedish National Space Board in Sweden. Additional support for science analysis during the operations phase is gratefully
acknowledged from the Istituto Nazionale di Astrofisica in Italy and the Centre
National d'\'Etudes Spatiales in France. This work performed in part under DOE
Contract DE-AC02-76SF00515. \\
We also acknowledge the Southern H-Alpha Sky Survey Atlas (SHASSA), which is supported by the National Science Foundation. }

\bibliographystyle{aa}
\bibliography{article}

\end{document}